\documentclass[%
reprint,
superscriptaddress,
 amsmath,amssymb,
 aps,
pra,
]{revtex4-2}

\usepackage{graphicx}
\usepackage{dcolumn}
\usepackage{bm}
\usepackage{hyperref}

\usepackage{cleveref}
\usepackage{xcolor,colortbl}
\usepackage{tcolorbox}
\usepackage{float} 
\usepackage{physics}

\newcommand{\ee}{\mathrm{e}}
\newcommand{\err}{{\sf Err}}

\newcommand{\Lc}{\Lambda^{(c)}}

\newcommand{\Mm}{M^{(m)}}

\usepackage[commandnameprefix=ifneeded]{changes}

\usepackage[resetlabels,labeled]{multibib}
\newcites{Math}{Math Readings}
\begin{document}

\preprint{APS/123-QED}


\title{Continuous variable measurement-device-independent quantum certification}


\author{B. L. Larsen}
\email{blula@dtu.dk}
\affiliation{Center for Macroscopic Quantum States (bigQ) Department of Physics, Technical University of Denmark, Fysikvej 307, 2800 Kongens Lyngby, Denmark}
\author{A. A. E. Hajomer}
\affiliation{Center for Macroscopic Quantum States (bigQ) Department of Physics, Technical University of Denmark, Fysikvej 307, 2800 Kongens Lyngby, Denmark}
\author{P. Abiuso}
\affiliation{Institute for Quantum Optics and Quantum Information - IQOQI Vienna,Austrian Academy of Sciences, Boltzmanngasse 3, A-1090 Vienna, Austria}
\author{S. Izumi}
\affiliation{Center for Macroscopic Quantum States (bigQ) Department of Physics, Technical University of Denmark, Fysikvej 307, 2800 Kongens Lyngby, Denmark}
\author{T. Gehring}
\affiliation{Center for Macroscopic Quantum States (bigQ) Department of Physics, Technical University of Denmark, Fysikvej 307, 2800 Kongens Lyngby, Denmark}
\author{J. S. Neergaard-Nielsen}
\affiliation{Center for Macroscopic Quantum States (bigQ) Department of Physics, Technical University of Denmark, Fysikvej 307, 2800 Kongens Lyngby, Denmark}
\author{A. Acín}
\affiliation{ICFO-Institut de Ciencies Fotoniques, The Barcelona Institute of Science and Technology,Av. Carl Friedrich Gauss 3, 08860 Castelldefels (Barcelona), Spain}
\author{U. L. Andersen}
\email{ulun@dtu.dk}
\affiliation{Center for Macroscopic Quantum States (bigQ) Department of Physics, Technical University of Denmark, Fysikvej 307, 2800 Kongens Lyngby, Denmark}

\date{\today}

\begin{abstract}
    Secure and reliable certification of quantum resources is a fundamental challenge in the advancement of next-generation quantum technologies, particularly as devices become more complex and integrated into practical applications, where parts of the system may be untrusted or inaccessible by users. Addressing this challenge requires certification methods that rely on minimal assumptions and limited trust while still faithfully and reliably verifying the quantum resources in question. For infinite-dimensional bosonic systems, existing certification methods rely on fully trusted and well-calibrated measurement systems, leaving significant security vulnerabilities open. In this work, we present the first experimental demonstration of measurement-device-independent (MDI) certification schemes for infinite-dimensional bosonic systems, where the certification process is conducted using entirely untrusted measurement devices and assuming only the trusted preparation of coherent states. Specifically, we implement schemes for the MDI certification of continuous-variable (CV) entanglement and the operation of an elementary optical CV quantum memory. 
    We leverage techniques of Bayesian metrology and
    exploit the practicality and accessibility of Gaussian quantum optics to achieve secure and efficient certification. These results demonstrate the potential of the MDI framework to enhance trust in quantum technologies with applications in quantum communication and quantum computing.

\end{abstract}

\maketitle

\section{Introduction}
\begin{figure}
    \includegraphics[width=\columnwidth]{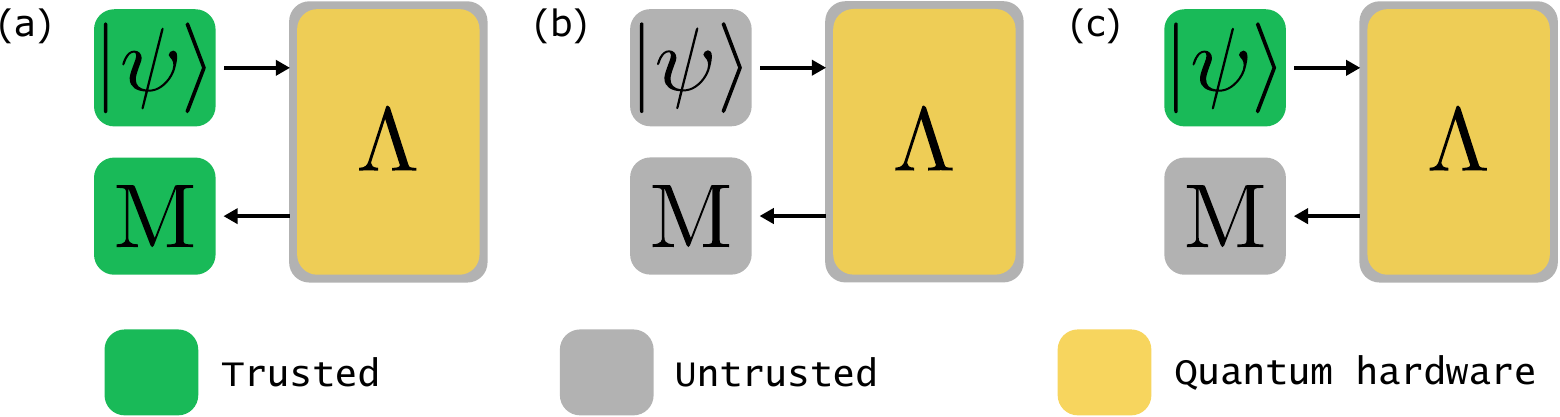}
    \caption{Conceptual diagrams for the certification of quantum hardware represented by a channel $\Lambda$. (a) Certification relying on a trusted (green) source of probe states $\ket{\psi}$ and a trusted measurement apparatus $\mathrm{M}$. (b) Certification in a fully device-independent scenario, where both the probe states and measurements are completely untrusted (grey). (c) Certification in the measurement-device-independent (MDI) framework, using trusted probe states but untrusted measurement devices to validate the quantum device.}
    \label{fig:concepts}
\end{figure}

How can a user with limited technological power certify the correct functioning of a complex uncharacterized quantum system? This is a crucial question in the development of quantum technologies, and various approaches have been explored to address it. A generic certification protocol begins by preparing the system in different possible states, $\{\psi^{(p)}\}$, then subjected to a set of possible channels $\{\Lc\}$ and finally performing measurements on the system $\{\Mm\}$. This process is repeated multiple times to gather sufficient data to estimate the statistics, $P_\text{obs}
    =\Tr[\Lc (\psi^{(p)})\Mm] \,$ from which relevant properties can be certified~\cite{Eisert2020,Kliesch2021}.  
Quantum state and quantum device tomography are well-known examples of this framework (see \cref{fig:concepts}a)~\cite{Lvovsky2009}. While these methods provide a robust and detailed form of certification, they are extremely resource-intensive, requiring not only complete knowledge of the implemented measurements but also precise calibration and control of the experimental setup.

Device-independent (DI) protocols take a drastically different approach: they make no assumptions about the specific forms of states, channels, or measurements, treating them as ``black boxes," with only their causal structure being given. For example, in a Bell test, 
if a Bell inequality violation is observed in the measurement statistics, the unknown state is confirmed to be entangled, regardless of the underlying hardware~\cite{Bancal2011,Baccari2017,Bowles2018}. 
Likewise, a range of protocols has been established for certifying quantum devices using a DI approach~\cite{brunner2014bell,Supic2020} (see \cref{fig:concepts}b). However, DI protocols are experimentally challenging and often require balancing the necessary assumptions with technological feasibility.

This  motivates the formulation of semi-device-independent approaches,  
particularly the measurement-device-independent (MDI) framework, in which only a set of prepared states is fully trusted while other elements remain uncharacterized. Buscemi~\cite{buscemi2012all} and Branciard \emph{et al.}~\cite{branciard2013measurement} introduced MDI entanglement witnessing protocols, proving that entanglement of any finite-dimensional state can be detected in a Bell experiment with trusted quantum inputs - a result that has been experimentally verified
\citep{nawareg2015experimental,xu2014implementation} and used in applications~\cite{kocsis2015experimental,guo2019experimental,zhao2020experimental}. Similar schemes have been developed and experimentally demonstrated for the MDI certification of quantum hardware (see \cref{fig:concepts}c)~\cite{pusey15quantum,rosset2018resource,mao2020experimentally,graffitti2020measurement,yu2021measurement}.

While MDI protocols have also been applied to continuous-variable (CV) quantum key distribution~\cite{pirandola2015high-rate}, all existing demonstrations of MDI resource and hardware certification have been restricted to finite-dimensional systems.
This is despite the fact that many critical quantum information protocols are based on CV systems~\cite{weedbrook2012gaussian}. These include CV quantum sensing~\cite{Tse2019,deAndrade2020}, CV quantum key distribution~\cite{zhang2024continuous}, and CV quantum computing~\cite{Zhong2020,Madsen2022}, each of which has exhibited quantum advantage. 

In this article, we introduce an adversarial-metrology framework that can be used for MDI resource certification in CV systems and encompasses the recent theoretical proposals~\cite{abiuso2021measurement,abiuso2023verification}. We present the first experimental MDI certification of CV entanglement, as well as MDI certification of an entanglement-preserving CV channel, akin to a CV transmission line or quantum memory. Specifically, leveraging the theoretical tools of quantum metrology, we implement CV witnesses for both state and channel certification that are independent of the specific operations of untrusted measurement devices, while relying only on the trusted preparation of coherent states.

\section{MDI certification via quantum metrology}

\begin{figure}
    \centering
    \centering
    \includegraphics[width=\columnwidth]{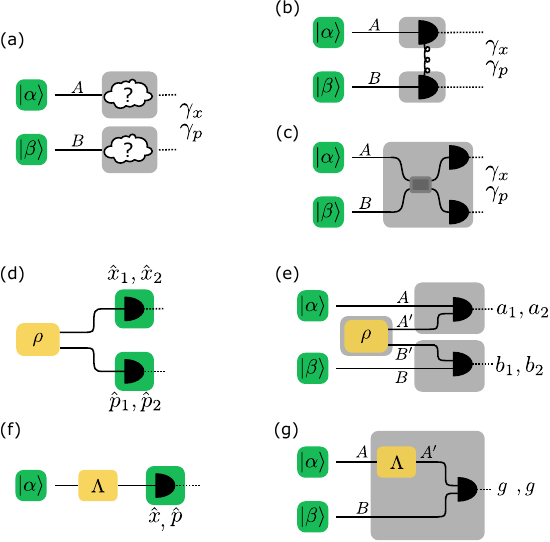}
    \caption{ Conceptual diagrams for MDI certification based on quantum metrology. {\sf(a)} Alice and Bob prepare trusted coherent states, $\{|\alpha = \alpha_x + i\alpha_p\rangle\}$ and $\{|\beta = \beta_x + i\beta_p\rangle\}$, which are sent to Eve. Eve is asked to estimate the joint variables, $\gamma_x = \alpha_x - \beta_x$ and $\gamma_p = \alpha_p + \beta_p$. Strategies restricted to local measurements assisted by at most one-way communication limit her estimation precision to the bound given by the inequality in~\cref{inequality}. However, by using either {\sf{(b)}} an entangled state or {\sf{(c)}} an entangling joint measurement strategy, this bound can be violated. {\sf(d-e)} Certification of entanglement using trusted measurements (left) versus MDI certification with untrusted measurements but trusted coherent state sources (right). The violation of bound~\eqref{inequality} implies that the shared state $\rho$ must be entangled. {\sf(f-g)} Certification of an entanglement-preserving channel, such as a quantum memory, using trusted measurements and sources (left) versus MDI certification with untrusted measurements and trusted coherent state sources (right). The violation of the bound in this case implies that the channel $\Lambda$ is not a measure and prepare process (hence it is EP).}
    \label{fig:MDI_metrology_and_witnesses}
\end{figure}

We start by considering the conventional strategy for certifying quantum resources, such as the detection of bipartite entanglement 
and the certification of an entanglement preserving quantum memory. 
In the former case, two parties, Alice and Bob, each receive one part of a bipartite state from an untrusted source, Eve. To verify the presence of CV entanglement, they perform well-calibrated quadrature measurement using homodyne or heterodyne detectors as illustrated in \cref{fig:MDI_metrology_and_witnesses}{\sf (d)}. The outcomes of these measurements are then evaluated using an entanglement witness such as the Duan-Simon inequality \cite{duan2000inseparability,simon2000peres} for separable states:
$
\langle \mathrm{EW} \rangle = \langle\Delta^2(\hat{x}_1-\hat{x}_2)\rangle+\langle\Delta^2(\hat{p}_1+\hat{p}_2)\rangle\geq2.
$
Here, using $\hbar = 1$, $\hat{x}_i$ and $\hat{p}_i$ represent the quadratures of the two modes, $i=1,2$ and $\langle \Delta^2 \hat{u} \rangle = \langle \hat{u}^2 \rangle - \langle \hat{u} \rangle^2$. If the measurement results of Alice and Bob lead to a violation of this inequality, the presence of entanglement is confirmed. However, this method critically relies on the accurate calibration of the homodyne detector systems. 
If the detectors are not properly calibrated, due to operational errors or external tampering, the computed variances may not accurately reflect the actual states. 
The measurement device must therefore be fully trusted. 

Another example is the certification of a quantum memory~\cite{rosset2018resource},
specifically corresponding to determining whether it preserves entanglement, which we dub entanglement preserving (EP) channel (as opposed to entanglement breaking channels~\cite{horodecki2003entanglement}). In other words, this involves verifying that the memory is able to store quantum states beyond simple classical storage, where states are measured, stored as classical bits, and then reconstructed~\footnote{That is, entanglement breaking (EB) channels are of the form~\cite{horodecki2003entanglement} $\Lambda^{\rm EB}[\rho]= \sum_a\; \rho^{(a)}\Tr[M^{a}\rho]$, where $M^{a}$ represents a measurement operator with outcome $a$, and $\rho^{(a)}$ the corresponding quantum state prepared. It follows that an entanglement breaking memory (transmission line) would only need to store (send) the classical data $a$ in order to simulate $\Lambda$ at arbitrary distance in time (space).}. Typically, this verification is performed by storing known quantum states, which are then released and subjected to quantum 
tomography to assess the fidelity of the memory process (as illustrated in \cref{fig:MDI_metrology_and_witnesses}{\sf (f)}. Also in this context, the reliability of the 
detectors is crucial; any malfunction or miscalibration can lead to inaccurate certification, thereby introducing security vulnerabilities. 

MDI protocols avoid these issues by not relying on the faithful functioning of the measurement apparatus.
In our case, we base the task of MDI entanglement or memory certification on a quantum metrology problem of estimating coherent states $|\alpha\rangle$ drawn from a Gaussian distribution
$P(\alpha)=(\pi\sigma_x \sigma_p)^{-1}\exp{-\frac{\alpha_x^2}{\sigma_x^2} -\frac{\alpha_p^2}{\sigma_p^2}}$, 
where $\alpha\equiv\alpha_x+i\alpha_p$. 
The minimum uncertainty in estimating the quadrature $\alpha_x = \langle \hat x/\sqrt{2} \rangle$ and $\alpha_p = \langle \hat p/\sqrt{2} \rangle$ 
of a coherent state from a Gaussian distribution can be derived using the techniques of Bayesian metrology. The resulting optimal bound for the summed uncertainties, governed by the a priori information of the distribution, is given by (see~\cite{yuen1973multiple,holevo2011probabilistic} and Supplementary Information)
\begin{align}
\label{eq:main_LB}
     \err(\alpha_x)+ \err(\alpha_p) &\geq \left(1+\frac{1}{2\sigma^{2}_x}+\frac{1}{2\sigma^{2}_p}\right)^{-1}=: v(\sigma_x,\sigma_p) \;,
\end{align}
where $\err(\theta):=\langle(\hat{\theta}-\theta)^2\rangle$ represents the average square error of any estimator $\hat{\theta}$ of the parameter $\theta$.
It is known that the optimal measurement strategy, saturating the bound for a symmetric Gaussian distribution ($\sigma_x=\sigma_p$), is balanced heterodyne detection, where the $\hat{x}$ and $\hat{p}$ quadratures are simultaneously measured with equal uncertainty~\cite{genoni2013optimal}. 

To derive a protocol for the MDI certification of quantum resources, we instead consider a variant of the previous metrology problem consisting in the estimation of the joint variables, $\gamma_x:=\alpha_x-\beta_x, \gamma_p:=\alpha_p+\beta_p$, from two independent coherent states, $|\alpha\rangle$ and $|\beta\rangle$, drawn from two Gaussian distributions. 
A straightforward approach is then to independently estimate the two coherent states with the minimal uncertainty found above, and subsequently estimate the joint variables through a linear combination of the measurement outcomes (see \cref{fig:MDI_metrology_and_witnesses} {\sf (a)}). This is indeed the optimal approach for estimating $\gamma_x$ and $\gamma_p$ when restricting the measurements to be local and assisted at most by one-way classical communication (see Supplementary Information). Under such restrictions, it follows from~\cref{eq:main_LB} that the uncertainty in estimating the joint parameters will be lower bounded by the inequality
\begin{align}
    \err (\gamma_x)+\err(\gamma_p) \geq {v(\sigma_{x,A},\sigma_{p,A})+v(\sigma_{x,B},\sigma_{p,B})}\;.
    \label{inequality}
\end{align}
However, this bound can in general be violated, 
for example when using shared entanglement between the two detection sites (relaxing the classical correlation restriction, see \cref{fig:MDI_metrology_and_witnesses}{\sf (b)}) or by performing non-local entangling measurements (thereby removing the locality restriction, see \cref{fig:MDI_metrology_and_witnesses}{\sf (c)})). These two types of entangling measurements are our primary tools for MDI resource certification: by broadcasting the outcomes of the entangling measurements, and subsequently violating~\cref{inequality}, 
we may verify the presence of entanglement or certify the function of a quantum device (cf. Methods). The verification procedure is completely independent of the measurement devices, only requiring that the coherent states received by the detection systems are fully trusted.

Having established the general MDI framework, we now specify it on the two certification tasks mentioned earlier: witnessing bipartite entanglement and certifying a quantum memory. The conceptual diagram of the corresponding entangling measurement schemes are illustrated in \cref{fig:MDI_metrology_and_witnesses} {\sf (e)} and {\sf (g)} and discussed in the following sections.

\section{MDI certification of CV entanglement}
\label{sec:res_EW}

\begin{figure*}[ht!]
    \centering
    \begin{minipage}{1\textwidth}
      \centering
      \includegraphics[width=\linewidth]{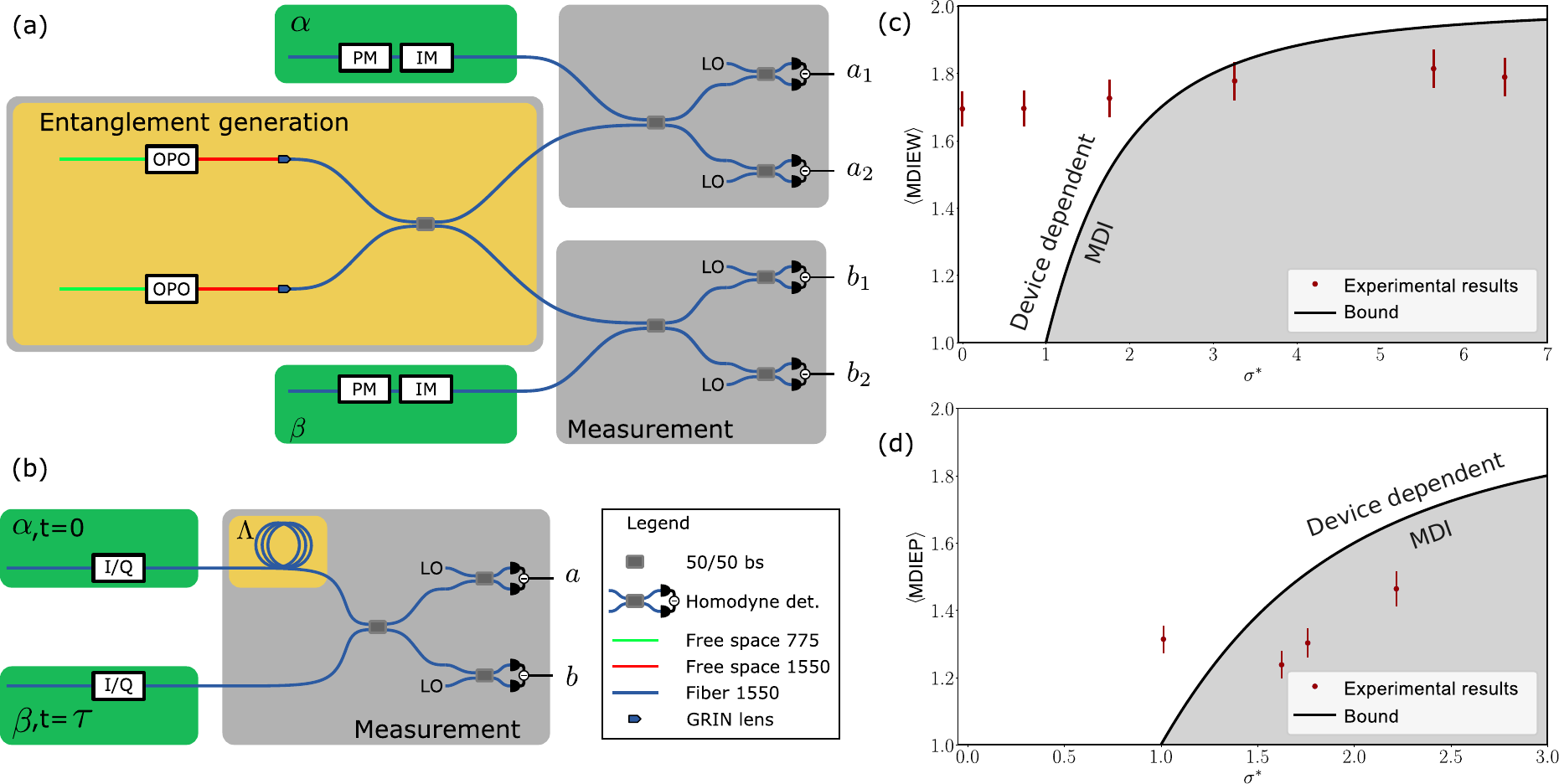}
    \end{minipage}
    \caption{MDI certification protocols of entanglement and a quantum memory.
    {\sf (a)} Schematic of setup for entanglement witness. For the experiment we generate an entangled state in the yellow section and perform the measurement of $\{\hat{x}, \hat{p} \}$ in each measurement boxes. All beam splitters are 50/50.
    {\sf (b)} Schematic of setup for memory verification. For the experiment we construct a memory for optical states with a 1 km long spool of fiber and perform the measurement of $\{\hat{x}, \hat{p} \}$ in the grey box. 
    Experimental results demonstrating the MDI witness of {\sf (c)} entangled TMSV state and {\sf (d)} EP quantum memory. 
    Red dots are the experimental values of the witness, while the solid black line indicates the theoretical upper bound of the witness. The shaded grey area delineates the MDI certification region for the protocols. The value of the witness in {\sf (c)} and {\sf (d)} is below 2 for all values of $\sigma^*$ thus violating the device dependent Simon-Duan bound and memory certification bound.}
    \label{fig:results}
\end{figure*}

We return to the scenario where Alice and Bob seek to verify the presence of bipartite CV entanglement provided by an untrusted source, Eve. 
This verification can be achieved through a violation of the coherent state estimation bound in \cref{inequality}, which crucially does not depend on the trustworthiness of the detectors (also provided by Eve). The detection strategy shown in \cref{fig:MDI_metrology_and_witnesses}{\sf (e)} allows for this. In this setup, Alice and Bob each prepare trusted coherent states of amplitudes $\alpha=\alpha_x+i\alpha_p$ and $\beta=\beta_x+i\beta_p$, sampled from Gaussian distributions with standard deviations $\sigma$. 
The two coherent states are then each measured locally, together with one mode of the bipartite entangled state under test. 
These two local measurements are each realized using two heterodyne detectors, effectively implementing a CV Bell state measurement. 
Notice here that this causal structure (see \cref{fig:MDI_metrology_and_witnesses}{\sf (e)}) is guaranteed if Alice and Bob operate locally the untrusted devices provided by Eve. 

Four quadrature measurements are therefore obtained
(coined $\{a_1,a_2,b_1,b_2\}$), 
and used to estimate $\gamma_{x,p}$. 
That is,
Alice and Bob compute the left-hand side of the witness inequality in \cref{inequality}, defined as $\langle\mathrm{MDIEW}\rangle= \langle (a_1 - b_1 - (\alpha_x-\beta_x))^2 \rangle+\langle (a_2 + b_2 - (\alpha_p+\beta_p))^2 \rangle \simeq \err (\gamma_x)+\err (\gamma_p)$, that is, taking $a_1 - b_1$ and $a_2 + b_2$ as the estimators of $\gamma_x$ and $\gamma_p$ respectively.
Entanglement is then confirmed if $\langle\mathrm{MDIEW}\rangle<2\nu(\sigma,\sigma)=2\sigma^2/(1+\sigma^2)$ assuming symmetric distributions $\sigma = \sigma_{x,A} = \sigma_{p,A} = \sigma_{x,B} = \sigma_{p,B}$ for simplicity.

To explore the parameter space for which entanglement can be detected, we compute the expected value of the left-hand side of the inequality in \cref{inequality} for $\rho$ being a two-mode squeezed vacuum state (TMSV) characterized by a squeezing parameter $r$ and subject to losses of $1-\eta$. The resulting value is $ \langle\mathrm{MDIEW}\rangle =2+\eta(e^{-2r}-1)$ (see Supplementary Information, or~\cite{abiuso2021measurement}). Thus, the condition for MDI entanglement detection is given by $\eta(1-e^{-2r})>2/(1+\sigma^2)$, assuming symmetric and identical coherent state distributions for both sources. It is evident that for large coherent state distributions ($\sigma \gg 1$), MDI detection of entanglement is feasible regardless of the degree of squeezing. In contrast, for narrow distributions, more stringent requirements are imposed on both the transmission coefficient, $\eta$, and the squeezing parameter, $r$. 

To demonstrate the MDI entanglement verification protocol (see~\cref{fig:results}), we generate bipartite entanglement in the form of a TMSV, achieved by interfering two single-mode squeezed states on a fiber beamsplitter. The resulting entangled modes are then sent to two separate measurement stations, where they are jointly measured with trusted coherent states using heterodyne detection systems. Each system comprises a symmetric beam splitter and two high efficiency balanced homodyne detectors, set to detect the conjugate quadratures, $x$ and $p$. The coherent states, used as trusted inputs, are generated at a sideband frequency of 3.1 MHz via an intensity and phase modulator (IM and PM) controlled by an arbitrary waveform generator.

The MDI protocol proceeds as follows:
First, full calibration of the input coherent state distributions is performed as a preliminary step. 
Following this, in each round, two trusted random coherent states are generated and injected, along with their respective TMSV entangled modes into the untrusted detection systems (see \cref{fig:results}{\sf (a)}). The measurement outcomes are then  recorded for subsequent analysis. The four outcomes, associated with the homodyne detectors, are then combined to form the joint variables $\{a_1-b_1,a_2+b_2\}$ used as estimators of $\{\gamma_x,\gamma_p\}$. We repeat the protocol with 1000 independently generated pairs of coherent states $\{\alpha,\beta\}$ drawn from the Gaussian distributions of widths $\sigma_{x,A}$, $\sigma_{p,A}$, $\sigma_{x,B}$ and $\sigma_{p,B}$. From these measurements, we compute the errors of the joint variables and sum the results to evaluate the left-hand side of \cref{inequality}, checking for any potential violations. To explore the effect of the input distributions, we vary the widths and, for each setting, perform measurements using 7000 copies of the same coherent states in order to estimate the statistical variance of the error of the measured witness. The summed errors obtained for six different distribution widths are shown in \cref{fig:results} {\sf (c)}, and compared against the threshold required for MDI verification of entanglement~\cref{inequality}. These results clearly demonstrate the successful observation of entanglement in a manner compliant with the MDI framework. Notice that here (and in the following memory certification \cref{sec:memory_cert})
we introduce in the figures an effective average variance, $\sigma^*$, defined such that $2 v(\sigma^*,\sigma^*)\equiv v(\sigma_{x,A},\sigma_{p,A})+v(\sigma_{x,B},\sigma_{p,B})$,
in order to account for the slight asymmetry in the widths.

It is important emphasizing that the metrological MDI witness~\cref{inequality} is fully adversarial. It means it accounts for the possibility that Eve, over time, could estimate the quantum input distribution (parametrized by $\sigma^*$ in our case) and consequently optimize her estimation strategy. However, the implemented experimental protocol is independent of the input distribution and is designed to achieve the same average error for any given value of $\gamma_{x,p}$. This shows that, for sufficiently large $\sigma^*$, the presence of entanglement can be reliable detected without requiring Eve to perform the (a priori nontrivial) estimation of the input distribution. 
In our theoretical analysis, we assumed that Alice and Bob could increase $\sigma^*$ at no added cost. In practice, however, the phase noise associated with high-amplitude coherent states hinders such assumption. Nevertheless, it is plausible that with knowledge of $\sigma^*$, the estimation protocol could be optimized to detect entanglement even at low values of $\sigma^*$. For instance, one such optimization involves post-processing the outputs of the homodyne measurements via a simple rescaling of the quadrature data: $\{a_1,a_2,b_1,b_2\} \rightarrow \{\epsilon a_1, \epsilon a_2, \epsilon b_1, \epsilon b_2\}$. 
In the Supplementary Information we show that such a rescaling extends the range of TMSV states for which entanglement is witnessed. The error observed by Alice and Bob with rescaling is: 
\begin{align}
    \nonumber
   \left\langle \mathcal{E} \right\rangle(\epsilon, \sigma^*, r) &= \err(\gamma_x) + \err(\gamma_p) \\
   &=  \epsilon^2 \ee^{-2r} + \epsilon^2 + 2 (\epsilon-1)^2 {\sigma^*}^2   \;,
\end{align}
which can violate~\cref{inequality} for $r>0$ and $\sigma^* > 0$. Importantly, any such optimization procedure can only \emph{improve} MDI entanglement detection, but cannot produce \emph{false} positives. That is, in the absence of entanglement ($r=0$), ~\cref{inequality} remains valid \emph{regardless} of the rescaling applied. 
%

In contrast, device-dependent entanglement witnessesmay yield false positives under similar conditions: For example, applying the Simon-Duan criterion to a rescaled system results in  $\langle \mathrm{EW} \rangle(\epsilon) = \langle \Delta^2 (\epsilon \hat{x}_1-\epsilon \hat{x}_2) \rangle +\langle \Delta^2 (\epsilon \hat{p}_1+\epsilon \hat{x}_2) \rangle = 2\epsilon^2$. Here a violation of the witness is observed for $\epsilon < 1 $, even in the absence of entanglement. 

\begin{figure}[ht!]
    \centering
    \includegraphics[width=\linewidth]{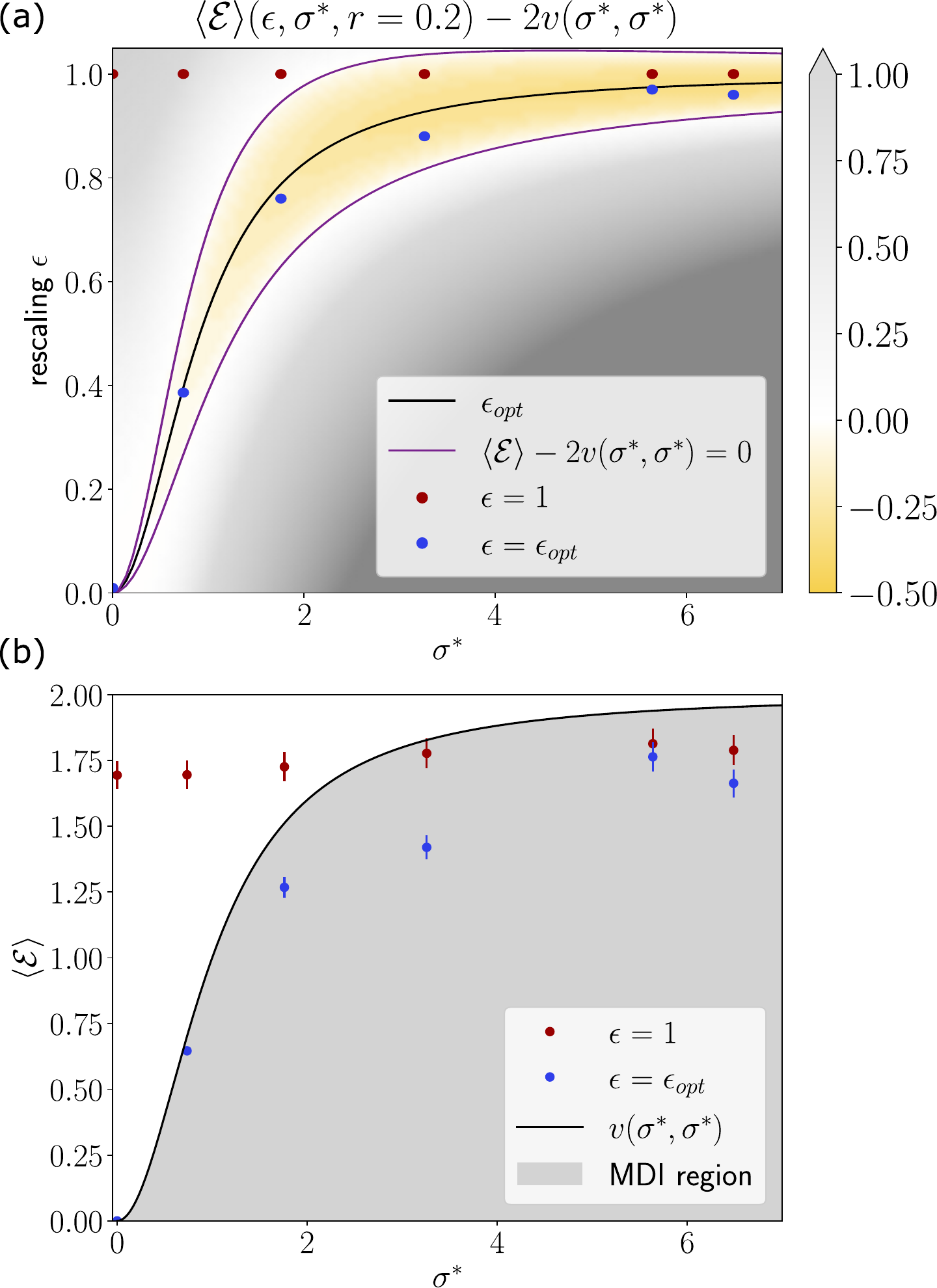}
    \caption{
     {\sf (a)} shows the value of $\left\langle \mathcal{E} \right\rangle(\epsilon, \sigma^*, r) - 2 v(\sigma^*,\sigma^*)$. A value below 0 implies MDI entanglement witness. The black curve is the optimal $\epsilon$ for the ideal case (no loss and phase noise). The white curves are $\left\langle \mathcal{E} \right\rangle(\epsilon, \sigma^*, r) = 2 v(\sigma^*,\sigma^*)$ between which the MDI entanglement witness can be performed. Red dots are experimental values without rescaling $\epsilon =1$ and Blue dots are with the optimal rescaling $\epsilon = \epsilon_{opt}$. {\sf (b)} is a reproduction of fig.~\ref{fig:results} {\sf (c)} but with optimized $\epsilon$. 
    }
    \label{fig:attack}
\end{figure}

We showcase this comparison in \cref{fig:attack} where we plot $\left\langle \mathcal{E} \right\rangle(\epsilon, \sigma^*, r) - 2 v(\sigma^*,\sigma^*)$, for which a value below 0 is a violation of the MDIEW witness, as a function of the rescaling $\epsilon$ and $\sigma^*$ with existing squeezing $r=0.2$. We show how the rescaling can lead to violation of the MDIEW bound with lower values of $\sigma^*$. 

\section{MDI certification of a CV quantum memory}
\label{sec:memory_cert}
Having successfully demonstrated MDI entanglement verification, we now employ the MDI framework to certify the operation of a quantum memory, ideally behaving as an identity channel ($\Lambda \approx \mathbb{I}$), in control of an untrusted party, Eve (see~\cref{fig:MDI_metrology_and_witnesses}{\sf(g)}). In this protocol the user, Alice, prepares trusted coherent states, $\ket{\alpha}$ and $\ket{\beta}$, sampled independently from Gaussian distributions. Unlike in the entanglement verification protocol, the states are prepared asynchronously: $\ket{\alpha}$ is stored in the quantum memory, while $\ket{\beta}$ is prepared later. After a storage period, the state $\ket{\alpha}$ is released from the memory and then jointly measured with $\ket{\beta}$ using a CV Bell state measurement. Eve performs the joint measurement and obtains the quadrature outcomes $\{g_x, g_p\}$. From these outcomes, we construct the witness:
$ \langle  \mathrm{MDIEP} \rangle= \langle (g_x - \gamma_x)^2 \rangle + \langle (g_p - \gamma_p)^2 \rangle \simeq \err( \gamma_x) + \err (\gamma_p)$. The entanglement-preserving nature of the quantum memory is verified if $ \langle \mathrm{MDIEP} \rangle< 2 v(\sigma^*,\sigma^*)$. In contrast to the previous protocol, where the violation indicated the presence of entanglement in the shared state, here the violation arises due to the local entangling measurement performed by Eve. This confirms the memory's capability to maintain quantum coherence, and in turn preserve entanglement when acting on part of an entangled state.

We now analyze the parameter space in which the EP nature of a quantum memory can be certified using the MDI framework. To do this, we compute the expected value of the left-hand side of ~\cref{inequality} for our system, taking into account both memory loss characterized by the transmission efficiency $\eta$ and the presence of excess noise $\xi$. Under these noise conditions, entanglement preservation is achieved when the following inequality holds: $ \langle \mathrm{MDIEP} \rangle= \frac{1}{2\eta} (2+\xi + \eta \xi)\geq 2 v(\sigma^*,\sigma^*)$ where we assume that the coherent state distributions are symmetric and identical (see Supplementary Information). The inequality is illustrated in \cref{fig:results}{\sf (d)}, showing the regions of the parameter space where the MDI entanglement-preserving  properties of the memory can be detected. The results indicate that only specific combinations of $\eta$, $\xi$ and $\sigma$, allow for the successful certification of EP features. For instance, in the regime of broad coherent state distributions ($\sigma\gg 0$) and no excess noise ($\xi=0$), MDI entanglement-preserving certification is feasible even with high memory loss. However, for narrow distributions and non-zero excess noise ($\xi\neq 0$), stricter requirements are imposed on the transmission efficiency $\eta$ for the detection of entanglement preservation.

We experimentally implement MDI memory certification using the setup shown in 
~\cref{fig:results}{\sf (b)}. The trusted coherent states,  $\ket{\alpha}$ and $\ket{\beta}$, are prepared in a 10 MHz wide frequency sideband using quadrature modulation. The quantum memory is implemented using a 1 km fiber spool, providing a storage time of 4.9 $\mu s$ and a transmission efficiency of $\eta \approx 0.8$ and negligible excess noise, $\xi \approx 0$. The CV Bell state measurement is performed by overlapping the two states on a polarization beam splitter, followed by a 90-degree polarization-based hybrid to detect the orthogonal quadrature. 
Additional details on the system implementation are provided in the Methods section and Supplementary Information.

Using the collected measurements, we calculate the errors of the joint variables following the procedure outlined in the entanglement witness protocol. We then evaluate $\langle \mathrm{MDIEP}\rangle$ to assess whether the obtained values violate the bound specified in~\cref{inequality}. To explore the impact of the input states, we systematically vary the widths of the coherent state distributions and perform measurements for each setting, with 2000 repetitions of 1000 coherent states. The results for four different distribution widths are presented in~\cref{fig:results}{\sf(d)}. These results clearly show violations of the MDI certification threshold for the three largest distribution widths, thereby confirming the successful and robust operation of the quantum memory.

\section{Conclusion}
In this work we have reported the successful certification of CV Gaussian entangled states and quantum memories within the MDI framework. Our approach relies solely on the trusted preparation of single-mode coherent states of light, arguably one of the simplest set of states to prepare in the lab. This minimal assumption is a significant advantage, as it allows the rest of the hardware to remain fully uncharacterized, offering strong security guarantees even when parts of the hardware are under the control of an untrusted party. Gaussian quantum optics, which has reached the level of a mature technology and is available in most experimental facilities, cannot be used for fully device-independent protocols. However, as demonstrated in this work, introducing minimal trust through the MDI scenario fundamentally changes the perspective. 

This demonstration of MDI resource and device certification represents an important step towards further applications of MDI protocols in the realm of CV quantum information. 
As quantum technologies continue to advance, we anticipate scenarios where users with limited technological capabilities will need to interact with large quantum service providers. In such contexts, MDI schemes like those demonstrated here could serve as essential building blocks for secure protocols and reliable certification, offering strong security guarantees even when the user must rely on external, potentially untrusted hardware.

\section*{Acknowledgments}
B.~L.~L., A.~A.~E.~H., S.~I., T.~G., J.~S.~N-N. and U.~L.~A. were supported by the Danish National Research Foundation (bigQ, DNRF0142), EU Horizon Europe (QSNP, grant no.\ 101114043) and the Innovation Foundation Denmark (CyberQ). In addition A.~A.~E.~H. and T.~G. were supported by Carlsberg Foundation (project CF21-046).

P.~A. was supported by the QuantERA II programme, that has received funding from the European Union’s Horizon 2020 research and innovation programme under Grant Agreement No 101017733, and from the Austrian Science Fund (FWF), project ESP2889224.

A.~A. was supported by the Government of Spain (Severo Ochoa CEX2019-000910-S, European Union NextGenerationEU PRTR-C17.I1 and FUNQIP), Fundació Cellex, Fundació Mir-Puig, Generalitat de Catalunya (CERCA program), the European Union (QSNP, 101114043 and Quantera project Veriqtas),
the ERC AdG CERQUTE and the AXA Chair in Quantum Information Science.

\section*{Methods}
\subsection{Units and normalization for continuous-variable quantum systems}
We use standard units for bosonic operators and quadratures, defined as
\begin{align}
    \hat{x}=\frac{\hat{a}+\hat{a}^\dagger}{\sqrt{2}}\;,\quad 
    \hat{p}= \frac{\hat{a}-\hat{a}^\dagger}{\sqrt{2}i}\;,
\end{align}
where $\hat{a}$, $\hat{a}^\dagger$ are the canonical annihilation and creation operators, respectively.
It follows that $[\hat{x},\hat{p}]=i$, and the uncertainty principle reads ${\rm Var}[\hat{p}]{\rm Var}[\hat{x}]\geq \frac{1}{4}$. Coherent states are defined as
\begin{align}
    \ket{\alpha}=e^{i \sqrt{2}\alpha_p \hat{x}-i\sqrt{2}\alpha_x\hat{p}}\ket{0}=e^{\alpha \hat{a}^{\dagger}-\alpha^*\hat{a}}\ket{0}\;,
\end{align}
where $\alpha=\alpha_x+ i\alpha_p$.
In these units, the variances of the quadratures for a coherent state (including the vacuum state) are equal to $\frac{1}{2}$, and the mean values satisfy $\bra{\alpha}\hat{x}\ket{\alpha}=\sqrt{2}\alpha_x$ and $\bra{\alpha}\hat{p}\ket{\alpha}=\sqrt{2}\alpha_p$.

\subsection{Witnessing of quantum resources via the limits of 1-LOCC measurements}
Our main tool for MDI certification is the inequality~\cref{inequality}, which is valid (see Supplementary Information) for 1-LOCC measurements (local operations and one-way classical communication).
A measurement on two systems $A$ and $B$ is 1-LOCC if it can be expressed as
of the form
\begin{align}
    \mathcal{M}^{{\rm(1-LOCC)}a,b}_{AB}=\sum_i P(i) \;\mathcal{M}^{(i)a}_A\otimes \mathcal{M}^{(i)b|a}_B\;,
    \label{eq:1-LOCC_meas}
\end{align}
where $\mathcal{M}^{(i)a}_A$ and $\mathcal{M}^{(i)b|a}_B$ are general local quantum measurements satisfying $\sum_a\mathcal{M}^{(i)a}_A=\mathbb{I}$ and $\sum_b\mathcal{M}^{(i)b|a}_B=\mathbb{I}$. $P(i)$ represents a random variable shared between the systems, and $\mathcal{M}^{(i)b|a}_B$ depends on $A$'s outcome $a$. This decomposition captures local measurements assisted by the possibility of one-way communication from $A$ to $B$.

\subsubsection{MDI entanglement witnessing}
In the Bell-like experiment presented in \cref{fig:MDI_metrology_and_witnesses} {\sf(e)}, coherent states $\ket{\alpha}$ and $\ket{\beta}$ are sent to causally-separated~\footnote{The causal separation can be taken as a reasonable experimental assumption, or it can be guaranteed via the requirement of the local measurements being space-like separated.} measurement stations $A$ and $B$, which may share a bipartite state $\rho_{A'B'}$. Given the locality of the two measurement stations $A$ and $B$, and making no other assumptions on the experimental devices other than this causal structure, the general form of their statistical output takes the form 
\begin{align}
\nonumber
    & P(a,b|\ket{\alpha},\ket{\beta}) \\ 
\nonumber    
    = &\Tr\left[(\ketbra{\alpha}_{A}\otimes \rho_{A'B'}\otimes \ketbra{\beta}_{B})M^{a}_{AA'}\otimes M^{b}_{BB'} \right] \\
    = & \Tr\left[(\ketbra{\alpha}_{A}\otimes \ketbra{\beta}_{B})\mathcal{M}^{a,b}_{AB}\right]\;,
    \label{eq:cal{M}_def}
\end{align}
where $\mathcal{M}_{AB}$ is an effective measurement acting on the input coherent states and producing outputs $\{a,b\}$. 
If the shared state $\rho_{A'B'}$ is separable, i.e. $\rho_{A'B'}=\sum_i p_i \rho_{A'}^{(i)}\otimes\rho_{B'}^{(i)}$, the induced measurement becomes separable:
\begin{align}
\nonumber
    \mathcal{M}^{a,b}_{AB} &:=\Tr_{A'B'}\left[\rho_{A'B'}\; M^{a}_{AA'}\otimes M^{b}_{BB'}\right]\\
    &\overset{\text{SEP}}{=}\sum_i p_i \; \mathcal{M}^{(i)a}_{A}\otimes \mathcal{M}^{(i)b}_{B}\;.
\end{align}
Such measurements are of the 1-LOCC form~\cref{eq:1-LOCC_meas}.
Thus, violation of the inequality~\cref{inequality} serves as an MDI witness of entanglement. 

Our experimental setup (\cref{sec:res_EW}) uses this framework to certify the entanglement of a TMSV state. Moreover, the related generalized MDI bound~\eqref{inequality} introduced in~\cite{abiuso2021measurement}, extends the Simon-Duan criterion~\cite{simon2000peres,duan2000inseparability} to detect any entangled Gaussian state. 

\subsubsection{MDI quantum memory certification}
The protocol for MDI quantum memory is schematically depicted in \cref{fig:MDI_metrology_and_witnesses} {\sf(g)}.
There, a random coherent state $\ket{\alpha}$ is sent to the adversary station at time $t=0$, while a second state $\ket{\beta}$ is sent after a time delay $t = \tau$, forcing the use of the memory for at least $\tau$. Under this causal structure, the output statistics are given by
\begin{flalign}
\label{eq:POVM_EB}
\nonumber
&\! P(b|\ket{\alpha},\ket{\beta}) 
\\
&\!=\Tr\left[M^{b}_{AB}(\Lambda\otimes\mathbb{I})[\ketbra{\alpha}_A \otimes\ketbra{\beta}_B]_{A'B}\right]
\\
&\!\overset{\text{EB}}{=}\sum_a \Tr\left[M^{b}_{A'B}(\rho^{(a)}_{A'}\otimes \ketbra{\beta}_{B})\right]\Tr\left[M^a_{A} \ketbra{\alpha}_{A}\right]\;.
    \label{eq:calM_EB}
\end{flalign}
where $M^{b}_{A'B}$ represents the measurement operator, and $\Lambda$ is the channel applied by the memory. The last equality~\eqref{eq:calM_EB} holds if the channel $\Lambda$ from $A$ to $A'$ is entanglement-breaking (EB). EB channels can always be represented in the measure-and-prepare form: $\Lambda[\rho_A]_{A'}\overset{\text{EB}}{=}\sum_a \rho^{(a)}_{A'}\Tr\left[M_A\rho_A\right]$~\cite{horodecki2003entanglement}.
In such case, without loss of generality, the virtual output $a$ can be copied and transmitted to the output port, resulting in an effective statistics $(a,b)$ obtained via the measurement
\begin{align}
    \mathcal{M}^{a,b}_{AB}  &\overset{\text{EB}}{=}M^{a}_{A}\otimes \Tr_{A'}\left[M^{b}_{A'B}\rho^{(a)}_{A'}\right]\;,
\end{align}
This measurement process adheres to the 1-LOCC form described in~\cref{eq:1-LOCC_meas}. Therefore, any violation of the inequality ~\cref{inequality} confirms that the memory is not entanglement-breaking, certifying its entanglement-preserving properties.

\subsection{Entanglement Witness}
\label{sec:MethodEW}

\subsubsection{Entangled resource}
The entangled TMSV state is generated by interfering two single-mode squeezed states produced by two identical optical parametric oscillators (OPO). Each OPO uses a bow-tie cavity configuration with a periodically poled potassium titanyl phosphate (PPKTP) crystal as the nonlinear medium. Each OPO is injected with a 1550 nm probe beam derived from a laser source (NKT Photonics X-15), while the pump at 775nm, required for the parametric process, is generated through second harmonic generation in another bow-tie cavity. To ensure stable operation, each cavity is locked using a dedicated locking beam at 1550 nm, which counter-propagates within the bow-tie cavity. The probe beam incorporates a piezo-mounted mirror, modulated with signals at $S_{OPO1} = 17$ kHz and $S_{OPO2} = 26$ kHz, respectively. For locking the relative phase of the OPOs, 1\% of the output light from the cavity is tapped off and demodulated to produce an error signal. This feedback signal allows the system to lock at 0 ($\pi$) relative phase between probe and pump, ensuring deamplification (amplification) of the parametric process. This results in one mode being squeezed in amplitude and the other in phase.

Feedback control for locking systems is managed using the PyRPL software \cite{Neuhaus2024a} running on RedPitaya FPGA boards. These boards feature integrated fast ADCs and DACs, enabling precise real-time control. The cavity locks are implemented using the Pound-Drever-Hall (PDH) technique~\citep{drever1983laser}. 

The single-mode squeezed states are coupled into optical fibers and subsequently interfered on a 50/50 fiber coupler with a controlled relative phase difference of $0$. This interference generates the TMSV state. To maintain the correct phase between the two single-mode squeezed states, a small fraction (1\%) of the output light from one arm of the fiber coupler is tapped using a 99/1 beam splitter. The relative phase is then measured, and a fiber stretcher is used to adjust the phase of one of the single-mode squeezed states accordingly. The phase locking system utilizes the modulation signal $S_{OPO2}$, which is also used for controlling the parametric gain of OPO2.

\subsubsection{Homodyne detection and data collection}
Quadrature measurements for both Alice and Bob are performed using four custom-built balanced homodyne detectors (HD), which measure the $\hat{x}$ and $\hat{p}$ quadratures. The phase between the local oscillator (LO) and the signal is locked using a fiber stretcher in the LO path. Locking ensures precise alignment with the $\hat{x}$ and $\hat{p}$ quadratures and is achieved through the sidebands $S_x$ and $S_p$ generated alongside the coherent states.

The HD employs high-efficiency photo-diodes with 99\% quantum efficiency (according to spec sheet from manufacturer). Light from the fiber is coupled onto the photo diodes using anti-reflection coated graded-index (GRIN) lenses. To address the slight imbalances in the 50/50 beam splitter (BS), caused by variations in its splitting ratio and polarization sensitivity, bending losses are introduced in one arm of the fiber arms. These losses are controlled using a small, adjustable 3D-printed component that bends the fiber, with fine-tuning performed via a screw mechanism. Additionally, a polarization controller is placed in the LO path to optimize the visibility between the LO and the signal, ensuring high-quality interference. 

The RedPitaya FPGA boards play a central role in the system. They generate intensity and phase modulation (IM and PM) signals and perform demodulation of the AC signals, measuring the locking signals $S_x$ and $S_p$ sidebands to maintain stable operation. 

The data from the homodyne detectors is collected using the RedPitaya boards. Demodulation of the AC signals is performed directly within the boards. To prevent interference and beating between the locking and data collection modulation signals, RedPitayas are synchronized through their clocks. The boards are connected to a PC via an Ethernet connection, which is used to store the measurement traces and generate the coherent state alphabets. To minimize laser noise during the measurement, a lock-measure scheme is implemented. In this scheme the lock-measure signal acts as a trigger, with the first half of the measurement trace taken during the locking phase and the second half during the measurement phase. To eliminate potential leakage in the acousto-optic modulator (AOM) during the transition, only the last 7700 data points of each measurement trace are used for analysis. A function generator is used to generate a square signal, used for both lock-measure and trigger of detection in the RedPiatayas. The square signal is operating at 10 Hz allowing sufficient time for data saving on the PC. Each individual measurement trace contains $2^{14}$ data points.


\subsection{Memory certification}
\label{sec:MethodMem}

The user (Alice) generated an ensemble of coherent states by modulating a 1550 nm continuous wave laser using in-phase and quadrature modulators driven by digital-to-analog converters (DACs). The two driving waveforms were generated digitally using the digital signal processing (DSP) chain, as described in~\citep{hajomer2023high}. This DSP module includes a quantum random generator (QRNG) based on vacuum fluctuations~\citep{hajomer2023high}, which provides the complex amplitude of the coherent states at a repetition rate of 10 MHz. These quantum symbols were then upsampled to the DACs sampling rate of 1 GSample/s and pulse-shaped using a root-raised cosine filter with a roll-off factor of 0.2. A 16 MHz pilot tone was added to the signal for phase compensation. Automatic bias controllers (ABC) were employed to stabilize the bias voltages of the IQ modulators. To fine-tune the size of the generated alphabet, variable optical attenuators (VOAs) were used after the IQ modulators.
Subsequently, one of the generated states was stored in the quantum memory, which was made of 1 km standard single mode fiber, with a storage time of 4.9~$\mu S$ and efficiency of $\approx 80\% $. To compensate for polarization rotation in the memory, a manual polarization controller was used before the memory. The other state was connected to the measurement device through a fiber patch cord. A polarization 90-degree hybrid relay was utilized to perform the CV Bell measurement~\citep{hajomer2023high}. Further details on the operational principles of this relay can be found in Ref~\citep{hajomer2023high}.

The signals were then detected and digitized using balanced detectors with a shot-noise limited bandwidth of 13 MHz and a 1 GSample/s analog-to-digital converter (ADC). Finally, the quadrature values were recovered using the DSP routine~\citep{hajomer2023high}, which includes phase compensation using the pilot tone and quadrature remapping, time synchronization, matched filtering, and downsampling. A figure of the experimental setup can be seen in Supplementary Information.

\bibliography{bibliography}
\newpage

\onecolumngrid
\newpage

\section*{Continuous variable measurement-device-independent quantum certification: Supplementary Information}

\appendix

\section{Main metrological bounds}

\subsection{Main bounds}
\label{sec:methods_main_bayesian_bound}
Here, we derive the Bayesian Cramér-Rao bound presented in ~\cref{eq:main_LB} of the main text. Specifically, we show that for a Gaussian prior distribution
\begin{align}
P(\alpha;\sigma_x,\sigma_p)=\frac{e^{-\alpha_x^2/\sigma^2_x-\alpha_p^2/\sigma^2_p}}{\pi\sigma_x\sigma_p}\;,
\label{eqapp:gaussian_prior}
\end{align}
the minimum average combined error for the simultaneous estimation of $\alpha_x$ and $\alpha_p$ encoded in the coherent state $\ket{\alpha_x+i\alpha_p}$, satisfies:
\begin{align}
    \err (\hat{\alpha}_{x})+\err(\hat{\alpha}_{p}) &\geq \left(1+\frac{1}{2\sigma^{2}_x}+\frac{1}{2\sigma^{2}_p}\right)^{-1}\;,
    \label{eqapp:CR_bayesian_bound}
\end{align}
where $\err(\theta):=\langle(\hat{\theta}-\theta)^2\rangle$ represents the average square error of any estimator $\hat{\theta}$ of the parameter $\theta$.
In the special case where $\sigma_x=\sigma_p$, the bound simplifies to $\err (\hat{\alpha}_{x})+\err(\hat{\alpha}_{p}) \geq \frac{\sigma^2}{1+\sigma^2}$. This specific result is known and has been previously reported in the literature (see e.g.~\cite{yuen1973multiple,holevo2011probabilistic,genoni2013optimal,demkowicz2015quantum,morelli2021bayesian}). Here, we generalize it to the case where $\sigma_x\neq\sigma_p$, following the approach of~\cite{yuen1973multiple}.
\paragraph*{Proof of~\cref{eqapp:CR_bayesian_bound} (\cref{eq:main_LB} in main text).}
We start by defining the complex estimator observable
\begin{align}
    \hat{\alpha}:=\hat{\alpha}_x+i\hat{\alpha}_p\;.
\end{align}
The goal is to find a lower bound on the average squared error:
\begin{align}
    \langle |\hat{\alpha}-\alpha|^2 \rangle=\int \dd^2\alpha P(\alpha) \Tr\left[(\hat{\alpha}-\alpha)(\hat{\alpha}^\dagger -\alpha^*)\ketbra{\alpha}\right]\;.
\end{align}
Consider the expression
\begin{align}
    B(\alpha):= P(\alpha)\Tr\left[(\hat{\alpha}-\alpha)\ketbra{\alpha}\right]\;.
    \label{eqapp:Balpha_def}
\end{align}
Here, $B(\alpha)$ represents the average bias of the estimator $\hat{\alpha}$ weighted by the prior distribution $P(\alpha)$, which is Gaussian in our case. Assuming the bias grows at most polynomially in $\alpha$, we have
\begin{align}
    \int\dd^2\alpha \frac{\partial B(\alpha)}{\partial \alpha} =0\;,
\end{align}
which translates to
\begin{align}
    \int\dd^2\alpha P(\alpha)\frac{\partial\ln P(\alpha)}{\partial \alpha}\Tr\left[(\hat{\alpha}-\alpha)\ketbra{\alpha}\right] + P(\alpha)\Tr\left[(\hat{\alpha}-\alpha)\ketbra{\alpha}L^\dagger_\alpha\right]=1\;,
\end{align}
where $L_\alpha$ is the \emph{Right Logarithmic Derivative} (RLD)~\cite{yuen1973multiple,genoni2013optimal}, defined by the equation $\partial_\alpha\rho(\alpha)=\rho(\alpha) L^\dagger_\alpha$.
We thus have
\begin{align}
    1=\int\dd^2\alpha P(\alpha)\Tr\left[(\hat{\alpha}-\alpha)(\frac{\partial \ln P(\alpha)}{\partial \alpha}+L_\alpha^\dagger)\right]\;.
\end{align}
Using the Cauchy-Schwarz inequality for the scalar product $\langle\!\langle O_1, O_2\rangle\!\rangle:=\int\dd^2\alpha P(\alpha)\Tr{\ketbra{\alpha}O_1^\dagger(\alpha) O_2(\alpha)}$, we obtain
\begin{align}
    1\leq \mathcal{F} \int \dd^2\alpha P(\alpha) \Tr\left[(\hat{\alpha}-\alpha)(\hat{\alpha}^\dagger -\alpha^*)\ketbra{\alpha}\right]\;,
\end{align}
or equivalently $\langle|\hat{\alpha}-\alpha|^2\rangle\geq \mathcal{F}^{-1}$, where $\mathcal{F}$ is defined as
\begin{align}
    \mathcal{F}:=\int \dd^2\alpha P(\alpha) \left[\ketbra{\alpha}\left(\frac{\partial \ln P(\alpha)}{\partial \alpha}^*+L_\alpha\right)\left(\frac{\partial\ln P(\alpha)}{\partial \alpha}+L_\alpha^\dagger\right)\right]\;.
\end{align}
Evaluating $\mathcal{F}$, we find that it separates into two terms: 
\begin{align}
    \mathcal{F}=\int \dd^2\alpha P(\alpha) \left|\frac{\partial \ln P(\alpha)}{\partial \alpha}\right|^2 +
    \int \dd^2\alpha P(\alpha) \left[\ketbra{\alpha} L_\alpha L_\alpha^\dagger \right] =: \mathcal{F}_P+\mathcal{F}_L \;.
\end{align}
The evaluation of $\mathcal{F}_L$ is standard but requires careful treatment due to the fact that $L_\alpha$ is ill-defined for pure states. The result of such procedure can be found e.g. in~\cite{yuen1973multiple,holevo2011probabilistic,genoni2013optimal}
and yields
\begin{align}
    \mathcal{F}_L=1\;.
\end{align}
For Gaussian prior distribution in Eq.~\eqref{eqapp:gaussian_prior}, we calculate 
\begin{align}
\nonumber
    \mathcal{F}_P &=\int \dd^2\alpha P(\alpha) \left|\frac{\partial \ln P(\alpha)}{\partial \alpha}\right|^2=\int \dd^2\alpha P(\alpha)\left|\frac{\partial}{\partial \alpha} (-\frac{\alpha_x^2}{\sigma_x^2}-\frac{\alpha_p^2}{\sigma_p^2})\right|^2\\
\nonumber
    &=\int \dd^2\alpha \frac{e^{-\frac{\alpha_x^2}{\sigma_x^2}-\frac{\alpha_p^2}{\sigma_p^2}}}{\pi\sigma_x\sigma_p}\left|\frac{\alpha_x}{\sigma_x^2}-i\frac{\alpha_p}{\sigma_p^2}\right|^2\\
    &=\frac{1}{2\sigma_x^2}+\frac{1}{2\sigma_p^2}\;
\end{align}
(notice that $\alpha$ and $\alpha*$ are independent variables and $\partial_\alpha(\alpha_x)=\partial_\alpha(\alpha_p)=\frac{1}{2}$).

Summing the above expressions for $\mathcal{F}_P$ and $\mathcal{F}_L$, one obtains finally
\begin{align}
    \langle|\hat{\alpha}-\alpha|^2\rangle\geq \left(1+\frac{1}{2\sigma_x^2}+\frac{1}{2\sigma_p^2}\right)^{-1}\;,
\end{align}
that is the starting bound~\eqref{eqapp:CR_bayesian_bound} used in the main text~\eqref{eq:main_LB}.

\subsection{1-LOCC reduction to local measurements}
\label{sec:methods_separable_povms}

Consider the situation in which two parameters $\alpha$ and $\beta$ are independently encoded in the product states
\begin{align}
    \rho(\alpha,\beta)=\rho(\alpha)_A\otimes \rho(\beta)_B\;.
\end{align}
Here we prove that the minimum estimation error for linear combinations of $\alpha$ and $\beta$, under the constraint that measurement is in the 1-LOCC class~\cref{eq:1-LOCC_meas}
\begin{align}
     \mathcal{M}^{{\rm(1-LOCC)}a,b}_{AB}=\sum_i p_i \;\mathcal{M}^{(i)a}_A\otimes \mathcal{M}^{(i)b|a}_B\;,
    \label{eqapp:1-LOCC_meas}
\end{align}
by the corresponding combination of minimum errors attained singularly on the two parameters independently, that is,
\begin{align}
    \min_{\mathcal{M}^{\rm 1-LOCC}_{AB}} \langle\left(f- (c_1\alpha+c_2\beta)\right)^2\rangle =
    c_1^2 \min_{\mathcal{M}_A} \langle(f_1- \alpha)^2\rangle
    + c_2^2 \min_{\mathcal{M}_B} \langle(f_2- \beta)^2\rangle\;.
\end{align}

\paragraph*{Proof.}
We explicitly consider the expression to minimize under the 1-LOCC constraint. The estimator $f$ is, in general, a function of the local measurement outputs in~\cref{eqapp:1-LOCC_meas} and possibly the index $i$. Thus, we have
\begin{align}
    \langle\left(f- (c_1\alpha+c_2\beta)\right)^2\rangle =\int P(\alpha)P(\beta)\sum_i p_i \sum_a\sum_b (f_i(a,b)-c_1\alpha+c_2\beta)^2 \Tr\left[(\mathcal{M}^{(i)a}_A\otimes \mathcal{M}^{(i)b|a}_B) (\rho(\alpha)_A\otimes \rho(\beta)_B)\right]\;.
\end{align}
The first simplification arises from the fact that this expression is an average over the index $i$, so it must be larger than its minimum addend. Hence, randomizing the $AB$-measurement does not reduce the error.
We can therefore restrict ourselves to
\begin{align}
    \langle\left(f- (c_1\alpha+c_2\beta)\right)^2\rangle \geq \int P(\alpha)P(\beta) \sum_a\sum_b (f(a,b)-c_1\alpha-c_2\beta)^2 \Tr\left[(\mathcal{M}^{a}_A\otimes \mathcal{M}^{b|a}_B) (\rho(\alpha)_A\otimes \rho(\beta)_B)\right]\;,
\end{align}
which simplifies to
\begin{align}
\label{eqapp:semic_f_exp}
    \langle\left(f- (c_1\alpha+c_2\beta)\right)^2\rangle \geq \sum\int P(a,\alpha)P(b,\beta|a) (f(a,b)-c_1\alpha-c_2\beta)^2\;,
\end{align}
where
\begin{align}
    P(a,\alpha) &= P(\alpha)\Tr\left[\mathcal{M}^{a}_A \rho(\alpha)_A\right]\;,\\
    P(b,\beta|a) &=  P(\beta)\Tr\left[\mathcal{M}^{b|a}_B \rho(\beta)_B\right]\;.
\end{align}
Next, we introduce now the quantity $\Bar{f}_\beta(a)$, which represents the average of $f$ over $\{b,\beta\}$  for fixed $\{a\}$:
\begin{align}
    \Bar{f}(a):=\sum_b\int\dd \beta\, P(b,\beta|a) f(a,b)\;.
\end{align}
Substituting this into the error expression, we can decompose it as 
\begin{align}
\nonumber
    &\sum\int P(a,\alpha)P(b,\beta|a) (f(a,b)-c_1\alpha-c_2\beta)^2\\=&
    \sum\int P(a,\alpha)P(b,\beta|a)
    \left((f(a,b)-\Bar{f}(a)+c_2\langle\beta\rangle-c_2\beta)^2+(\Bar{f}(a)-c_1\alpha-c_2\langle\beta\rangle)^2\right)\;,
    \label{eqapp:alm_the}
\end{align}
This decomposition is achieved by adding and subtracting $\Bar{f}(a)-c_2\langle\beta\rangle$ and recognizing that the cross term averages to zero. Specifically, the cross term $\propto (f(a,b)-\Bar{f}(a)+c_2\langle\beta\rangle-c_2\beta)(\Bar{f}(a)-c_1\alpha-c_2\langle\beta\rangle)$ vanishes due to the successive averaging over $b$, $\alpha$ and $a$. Explicitly, we have
\begin{align}
    \sum_a\int \dd\alpha\, P(a,\alpha) (\Bar{f}(a)-c_1\alpha-c_2\langle\beta\rangle)\equiv \sum_{a,b}\int \dd\alpha\dd\beta P(a,\alpha)P(b,\beta|\alpha) (f(a,b)-c_1\alpha-c_2\beta)=0\;,
\end{align}
where the second equality holds because $f(a,b)$ can always be assumed to be globally unbiased. This assumption is justified since the average squared error is minimized for globally unbiased estimators.
We now define two separate estimators \cref{eqapp:alm_the}: 
\begin{align}
    f_1(a):=&\Bar{f}(a)-c_2\langle\beta\rangle\;,
    \\
    f_2(b):=& f(\Tilde{a},b)-\Bar{f}(\tilde{a})+c_2\langle\beta\rangle\;, \\
    \nonumber 
\end{align}
where $\tilde{a}$ is the argument minimizing the residual error:
\begin{align}
\Tilde{a}=&\text{argmin} \sum_b\int\dd\beta\, P(b,\beta|a)
    (f(a,b)-\Bar{f}(a)+c_2\langle\beta\rangle-c_2\beta)^2\;.    
\end{align}
Finally, connecting \cref{eqapp:semic_f_exp} with \cref{eqapp:alm_the}, with the above definitions, the error can be expressed as
\begin{align}
    \langle(f-(c_1\alpha+c_2\beta))^2\rangle \geq \langle(f_1-c_1\alpha)^2+(f_2-c_2\beta)^2\rangle
\end{align}
This confirms that the minimum error is dictated by the independent estimation errors of $\alpha$ and $\beta$, weighted by the coefficients $c_2^2$ anbd $c_2^2$, respectively, as desired.

\newpage 
\section{Experimental details}

\begin{figure}
    \centering
    \includegraphics[width=\columnwidth]{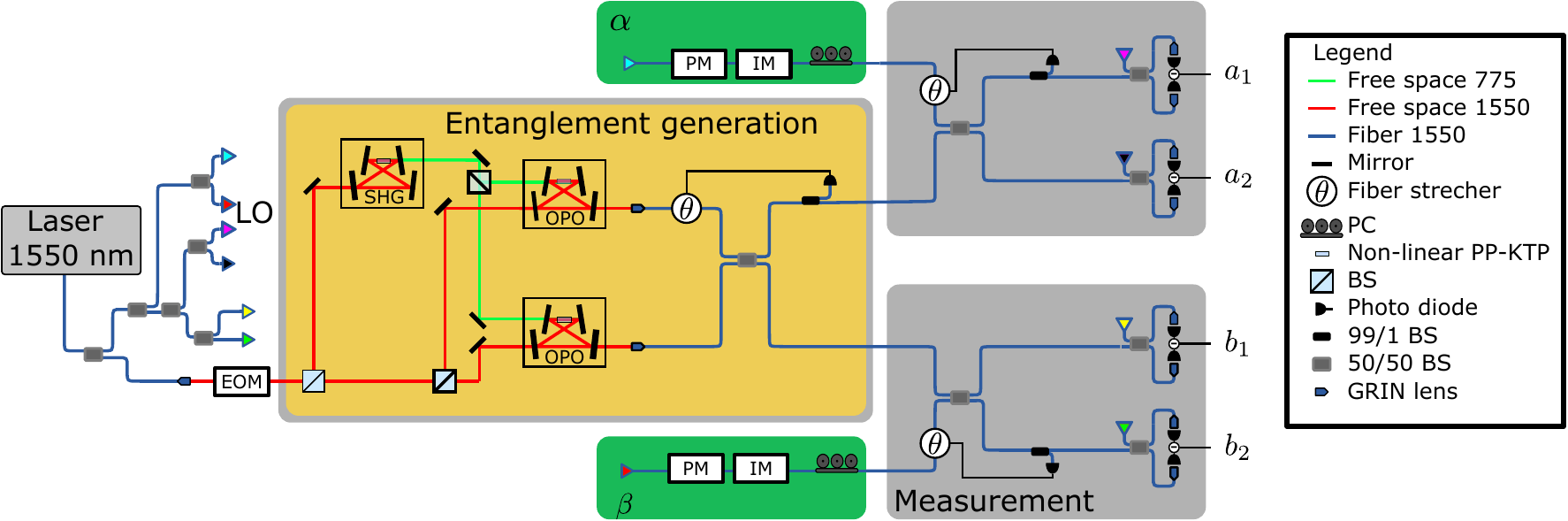}
    \caption{Detailed schematic of the setup for the generation of entangled two-mode squeezed vacuum state (TMSV), coherent state generation and measurement.}
    \label{fig:setup2}
\end{figure}

\begin{figure}
    \centering
    \includegraphics[width=\columnwidth]{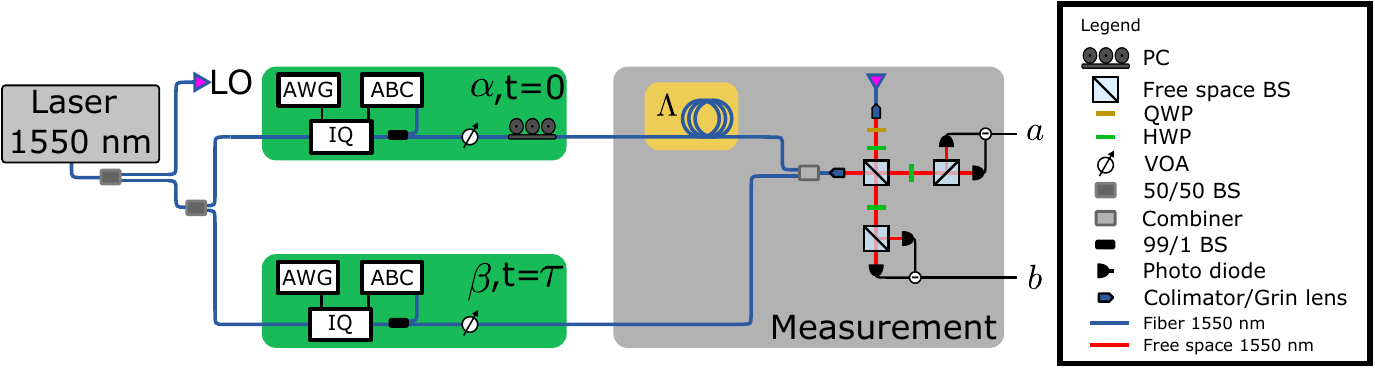}
    \caption{Experimental setup for quantum memory certification.}
    \label{fig:setup3}
\end{figure}

Detailed schematics of the setup can be seen for Entanglement witness in  \cref{fig:setup2} and for memory certification in \cref{fig:setup3}.

\subsection{Entanglement witness: Coherent state generation}
To calibrate the relationship between the amplitude of the resulting amplitude of generated coherent states, we varied the modulation voltage while measuring only the generated coherent states. A clear linear relationship was observed (\cref{fig:linscale}), allowing Alice and Bob to accurately generate known coherent states for the trusted input.

Coherent states for both Alice and Bob were generated using the same method. IM and PM from "iXblue" were used to modulate $\alpha_x$ and $\alpha_p$ components of the states, respectively. The coherent states were generated at a sideband frequency of $S_{meas} = 3.1$ MHz. By varying the modulation voltage, coherent states with different amplitudes were generated in the sideband. In addition, the IM and PM generated sidebands at $S_x = 31$ and $S_p = 41$ MHz, respectively, which were used to lock the homodyne detectors to the $x$- and $p$-quadratures. The RedPitaya boards running PyRPL software were used for generating the states and locking signals. 

\begin{figure}
\centering
\begin{minipage}{.45\columnwidth}
  \centering
  \includegraphics[width=0.95\linewidth]{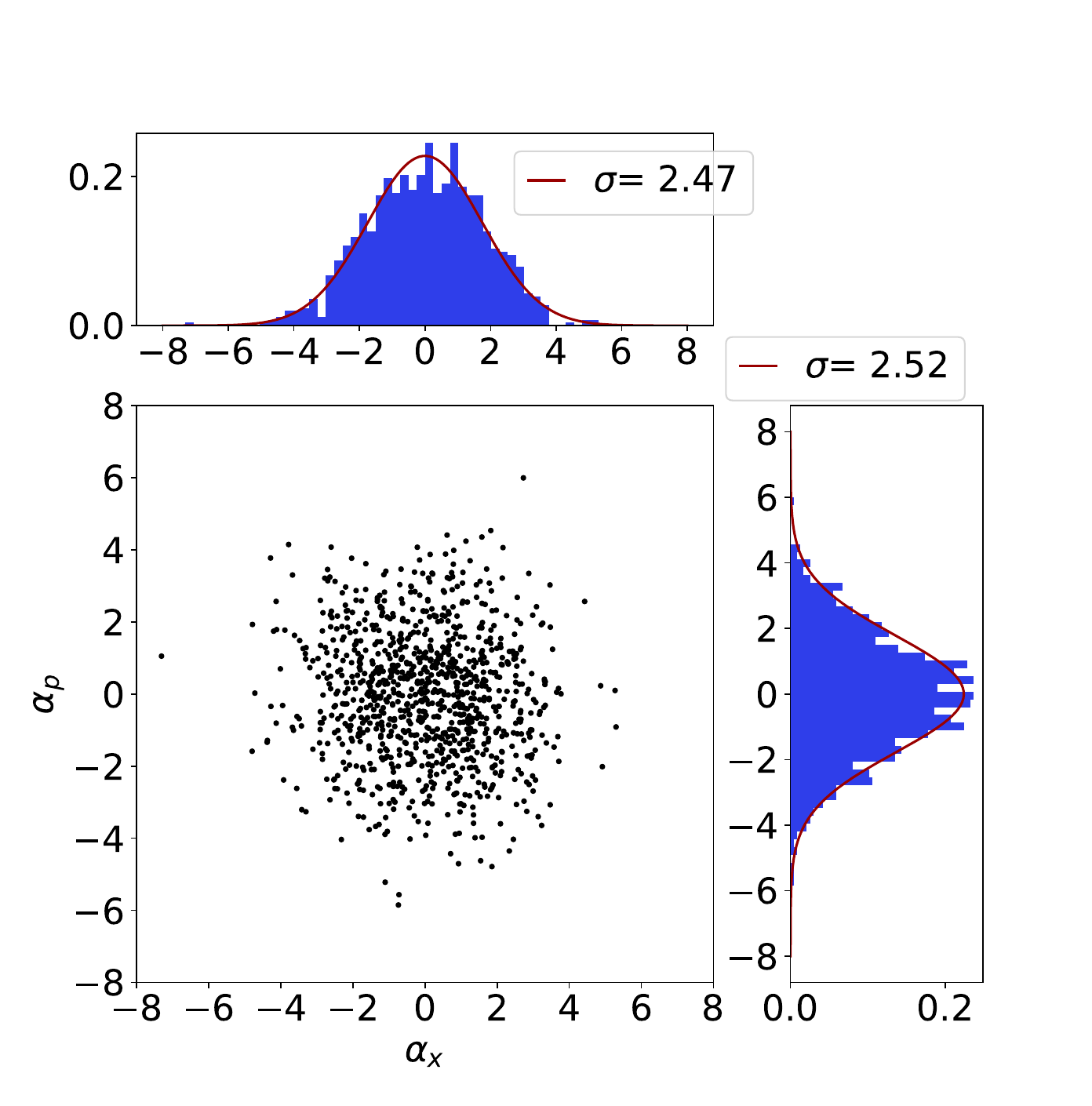}
\end{minipage}%
\begin{minipage}{.45\columnwidth}
  \centering
  \includegraphics[width=0.95\linewidth]{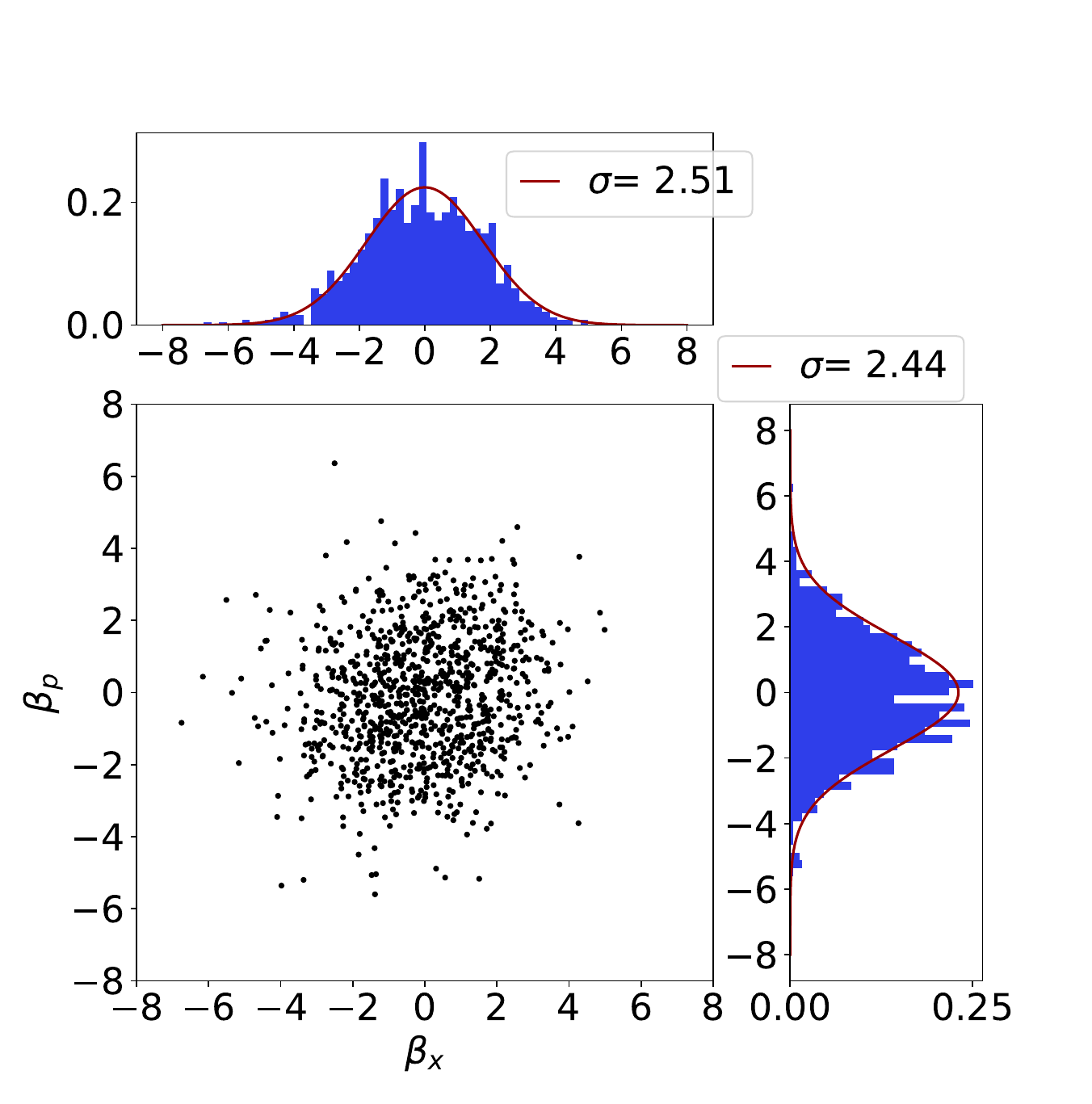}
\end{minipage}
    \caption{Scatter plots and histograms of the $\hat{x}$- and $\hat{p}$- quadrature for the 1000 states generated with $\sigma \approx 2.5$. Left: Alice's states. Right: Bob's states.}
\label{fig:stateplots}
\end{figure}

\begin{figure}
    \centering
    \begin{minipage}{.85\columnwidth}
      \centering
      \includegraphics[width=0.95\linewidth]{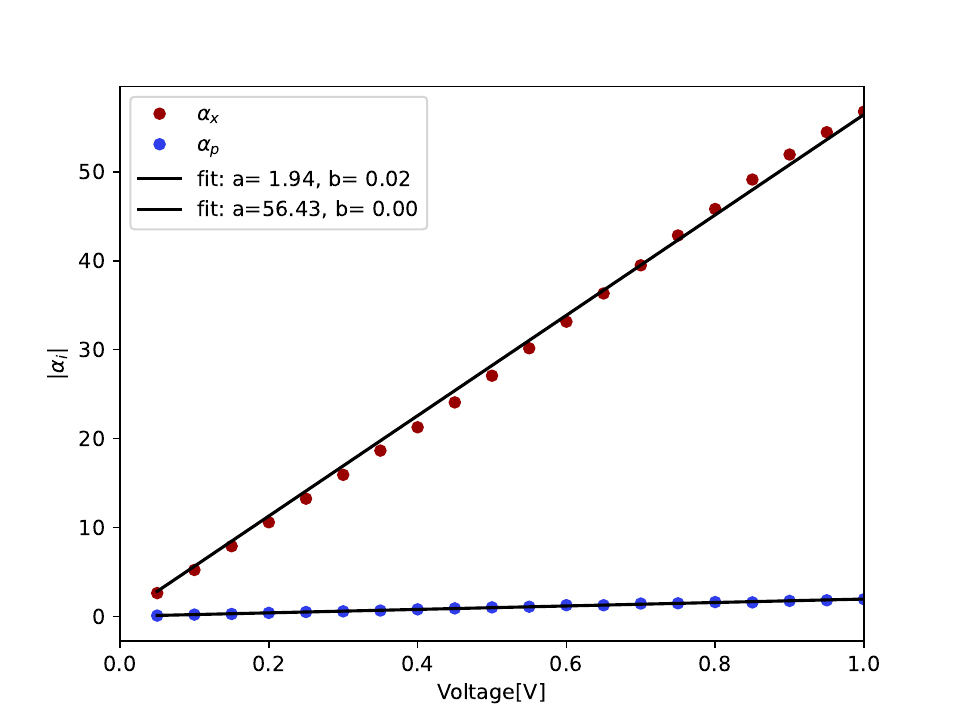}
    \end{minipage}
    \caption{Linear relationship between the modulation voltage applied to the phase modulator (PM) and the amplitude of the generated state in the $\hat p$-quadrature. Additionally, a linear dependence is observed in the $\hat x$-quadrature, indicating residual modulation. This behavior highlights the need for compensation when generating coherent states with precise quadrature control.} 
    \label{fig:linscale}
\end{figure}

For the experimental alphabet, 1000 coherent states were randomly generated from a Gaussian distribution. A sample of these states is shown in \cref{fig:stateplots}, with scatter plots and histograms illustrating the x- and p-quadratures for Alice and Bob.

The phase between the coherent states and the TMSV was locked after interference at a 50/50 BS. This was achieved using a 99/1 BS in one arm to monitor the phase, and a fiber stretcher before the 50/50 BS to apply corrections. 

During calibration, we found that generating a coherent state using only the IM, intended to produce amplitude modulation in the $\hat{x}$-quadrature, also introduced a small residual modulation in the $\hat{p}$-quadrature, and vice versa. This effect is likely due to residual amplitude modulation caused by the crystal's response to the modulation frequency~\citep{dominguez2017fundamental}. Fortunately, this response scales linearly with the amplitude modulation (see \cref{fig:linscale}), allowing us to account for it when generating the coherent state alphabet. By applying a compensating modulation through the PM, we effectively canceled the unintended modulation in $\hat{p}$ from the IM, enabling independent generation of any desired coherent state. 

\subsection{Entanglement witness: Efficiency budget}
In this section, we analyze the efficiency of the setup, focusing on state generation, transmission of the state to the detectors and the detection efficiency. 
For state generation, the escape efficiencies of the optical parametric oscillators (OPOs) were measured as $97.5 \pm 0.1 \%$ (intracavity loss: $0.255\%$) for OPO 1, and $94.4 \pm 0.1 \%$ (intracavity loss $0.588\%$) for OPO 2. The coupling efficiencies of the OPO outputs into the fibers were $87 \pm 1 \%$ and $84 \pm 1 \%$, respectively.

 The optical transmission in the fiber components includes losses from two 50/50 fiber couplers (used for generating the EPR state and mixing the coherent states), a 99/1 fiber coupler for tapping light for locking, fiber splicing and fiber absorption. The total transmission efficiencies to the 4 homodyne detectors were measured as 20 \%, 22 \%, 21 \% and21 \% for homodynes 1,2,3 and 4, respectively.

The detector efficiency of each homodyne detector is determined by the diode efficiency (99\%, form spec sheet), visibility (optimized through fiber coupling for spatial mode matching), and electronic noise clearance (98-99 \% with a signal-to-noise ratio of 18-20 dB). The resulting detection efficiencies were 95 \%, 96 \%,  94 \% and 95 \% for homodynes 1,2,3 and 4, respectively. 

Using the efficiencies from state generation, transmission, and detection, we calculated the expected squeezing levels. For the total efficiency $\eta$, the bandwidth of the cavity $\Omega = 8.35$ MHz, the sideband frequency of the measurement $\Omega = 3.1$ MHz, and $\zeta = \Omega \sqrt{P/P_{thr}}$, where $\frac{P}{P_{thr}} \approx 0.40$ (with $P$ and $P_{thr}$ being the pump and threshold power), we find 
\begin{align*}
    \langle \Delta^2(\hat{x}_1 - \hat{x}_2) \rangle &= \langle \Delta^2(\hat{p}_1 + \hat{p}_2) \rangle = \frac{1}{2} - 2 \eta \frac{\Omega \zeta}{(\Omega + \zeta)^2 + \omega^2} \approx 0.42 \\
    \langle \Delta^2(\hat{x}_1 + \hat{x}_2) \rangle &= \langle \Delta^2(\hat{p}_1 - \hat{p}_2) \rangle = \frac{1}{2} + 2 \eta \frac{\Omega \zeta}{(\Omega + \zeta)^2 + \omega^2} = 1.38
\end{align*}

For measurements with $\sigma=0$, the observed squeezing levels were
\begin{align*}
    \langle \Delta^2(\hat{x}_1 - \hat{x}_2) \rangle &= 0.428 \pm 0.019 \\
    \langle \Delta^2(\hat{x}_1 + \hat{x}_2) \rangle &= 1.206 \pm 0.068 \\
    \langle \Delta^2(\hat{p}_1 + \hat{p}_2) \rangle &= 0.417 \pm 0.019 \\
    \langle \Delta^2(\hat{p}_1 - \hat{p}_2) \rangle &= 1.429 \pm 0.063
\end{align*}

The measured squeezing was slightly lower than the expected, while the anti-squeezing was slightly higher. Introducing an additional $20 \%$ loss in the calculations aligns the expected squeezing with the observed values, giving $ S = 0.43$ for squeezing and and $AS = 1.20$ for anti-squeezing.  This suggests the presence of an unknown loss in the system.

\subsection{Squeezing-loss analysis for TMSV}
We analyze the proposed experimental protocol for MDI entanglement witnessing, schematically depicted in \cref{fig:blackboc}. By taking the outputs $a_1-b_1$ and $a_2+b_2$ as estimators for $\alpha_x-\beta_x$ and $\alpha_p+\beta_p$, respectively, the MDI witness value from~\cref{inequality} becomes 
\begin{align}
\langle\mathrm{MDIEW}\rangle &= \langle (a_1 - b_1 - (\alpha_x-\beta_x))^2 \rangle+\langle (a_2 + b_2 - (\alpha_p+\beta_p))^2 \rangle\\
&=\left\langle \left(\frac{\hat{x}_A+\hat{x}_a-\hat{x}_B-\hat{x}_b}{\sqrt{2}}-(\alpha_x-\beta_x)\right)^2 \right\rangle 
+ \left\langle \left(\frac{\hat{p}_A-\hat{p}_a+\hat{p}_B-\hat{p}_b}{\sqrt{2}}-(\alpha_p+\beta_p)\right)^2 \right\rangle
\end{align}
Now, consider the source $\rho_{AB}$ as a two-mode squeezed vacuum (TMSV). Using the properties of TMSV, we know that $\langle\hat{x}_A\rangle=\langle\hat{p}_A\rangle=\langle\hat{x}_B\rangle=\langle\hat{p}_B\rangle=0$, as well as $\langle\hat{x}_a\rangle=\sqrt{2}\alpha_x$, $\langle\hat{x}_b\rangle=\sqrt{2}\beta_x$ (and similarly for $\hat{p}$). Substituting these into the above expression, we simplify:
\begin{align}
    \langle\mathrm{MDIEW}\rangle &=\frac{1}{2}\left(\langle(\hat{x}_A-\hat{x}_B)^2\rangle+\Delta\hat{x}_a^2+\Delta\hat{x}_b^2+\langle(\hat{p}_A+\hat{p}_B)^2\rangle+\Delta\hat{p}_a^2+\Delta\hat{p}_b^2\right)\\
    &=1+\frac{1}{2}\left(\langle(\hat{x}_A-\hat{x}_B)^2\rangle+\langle(\hat{p}_A+\hat{p}_B)^2\rangle\right)
\end{align}
For a TMSV with squeezing parameter $r$, we have  $\langle(\hat{x}_A-\hat{x}_B)^2\rangle+\langle(\hat{p}_A+\hat{p}_B)^2\rangle=2e^{-2r}$. Substituting this result, the MDI witness evaluates to $\langle\mathrm{MDIEW}\rangle=1+e^{-2r}$, which is strictly less than 2 for any nonzero squeezing $r>0$, confirming the presence of entanglement.

Optical losses in the TMSV can be modeled as a beam splitter with transmissivity $\eta$. The modes transform as 
\begin{align*}
    \{\hat{x}_A,\hat{p}_A,\hat{x}_B,\hat{p}_B\}\rightarrow
    \sqrt{\eta}\{\hat{x}_A,\hat{p}_A,\hat{x}_B,\hat{p}_B\}+\sqrt{1-\eta}\{\hat{x}_{V_A},-\hat{p}_{V_A},\hat{x}_{V_B},-\hat{p}_{V_B}\}
\end{align*}
where $V_A$ and $V_B$ are vacuum modes.
Including such losses, the MDI witness becomes
\begin{align}
    \langle\mathrm{MDIEW}\rangle=1+\eta e^{-2r}+1-\eta=2+\eta(e^{-2r}-1).
\end{align}
This expression shows that losses reduce the effectiveness of the witness. 
We plot the resulting value of $ \langle \mathrm{MDIEW} \rangle$ as a function of $\eta$ and $r$ in \cref{fig:Countous} {\sf (a)}. The plot illustrates the critical interplay between squeezing and losses, demonstrating that the witness remains below the threshold for sufficiently high $\eta$ and $r>0$.

\subsection{Rescaling of measurements by Eve}

A rescaling of measurements, either due to experimental errors or intenational tampering by Eve, can lead to false violation of the Simon-Duan criterion. For example, a separable vacuum state can erroneously violate this criterion when the measurements are multiplied by a rescaling factor $\epsilon <1 $. 
\begin{align*}
    \langle \mathrm{EW} \rangle = \langle \Delta^2 (\hat{x}_1-\hat{x}_2) \rangle +\langle \Delta^2 (\hat{p}_1+\hat{x}_2) \rangle = 2  \rightarrow \\
    \langle \mathrm{EW} \rangle(\epsilon) = \langle \Delta^2 (\epsilon \hat{x}_1-\epsilon \hat{x}_2) \rangle +\langle \Delta^2 (\epsilon \hat{p}_1+\epsilon \hat{x}_2) \rangle = 2\epsilon^2 < 2 
\end{align*}

\begin{figure}[ht!]
    \centering
    \includegraphics[width=0.7\linewidth]{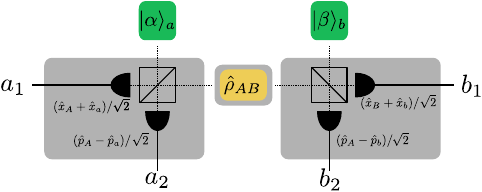}
    \caption{Eve's measurement setup. In the rescaling attack, Eve multiplies here measurement outcomes $\{a_1,a_2,b_1,b_2\}$ with the rescaling factor $\epsilon$. The MDI witness ensures that only if the state $\hat{\rho}_{AB}$ is entangled can Eve violate the bound in \cref{inequality}, even with rescaling.}
    \label{fig:blackboc}
\end{figure}

For the MDI protocol, Eve estimates the coherent states sent by Alice and Bob by mixing the states $\hat{x}_a$ with her state $\hat{x}_A$ using a 50/50 BS and measuring the $\hat{x}$- and $\hat{p}$-quadratures of the outputs (See figure \cref{fig:blackboc}). For a rescaling attack, Eve multiplies her measurement outcomes $\{a_1,a_2,b_1,b_2\}$ by the factor $\epsilon$.  

With this setup, with a rescaling by $\epsilon$, the error ($\err(\gamma_x)$) becomes:
\begin{align}
    &\left\langle \mathcal{E} \right\rangle(\epsilon)_x \equiv \left\langle \left(\frac{\epsilon}{\sqrt{2}} (\hat{x}_a + \hat{x}_A - \hat{x}_b - \hat{x}_B) - (\alpha_x - \beta_x)\right)^2 \right\rangle  \\
    & \underbrace{=}_{\langle \hat{x}_A \rangle = \langle \hat{x}_B \rangle = 0} \frac{\epsilon^2}{2} \left\langle  \hat{x}^2_a +  \hat{x}^2_A +  \hat{x}^2_b +  \hat{x}^2_B \right\rangle 
    + \alpha_x^2 + \beta_x^2 - \epsilon^2\langle\hat{x}_A \hat{x}_B\rangle \\
    & - \epsilon^2 \langle\hat{x}_a\rangle\langle\hat{x}_b\rangle - \sqrt{2}\epsilon \alpha_x \langle \hat{x}_a \rangle + \sqrt{2}\epsilon \alpha_x \langle \hat{x}_b \rangle 
    + \sqrt{2}\epsilon \beta_x \langle \hat{x}_a \rangle \\
    & - \sqrt{2}\epsilon \beta_x\langle\hat{x}_b\rangle  + 2 \alpha_x \beta_x.
    \label{eq:EpsilonModel}
\end{align}

Using $\langle \hat{x}_a \rangle = \sqrt{2} \alpha_x$ and $\langle \hat{x}_b \rangle = \sqrt{2} \beta_x$, we find 
\begin{align*}
    &\left\langle \mathcal{E} \right\rangle(\epsilon)_x = \left\langle \left(\frac{\epsilon}{\sqrt{2}} (\hat{x}_a + \hat{x}_A - \hat{x}_b - \hat{x}_B) - (\alpha_x - \beta_x)\right)^2 \right\rangle =  \frac{\epsilon^2}{2} \langle (\hat{x}_A-\hat{x}_B)^2 \rangle \\
    &+ \epsilon^2 \left(\alpha_x^2 + \beta_x^2+\frac{1}{2}\right) + \alpha_x^2 + \beta_x^2 - 2\epsilon^2 \beta_x\alpha_x + 2\epsilon(2 \alpha_x\beta_x-\alpha_x^2-\beta_x^2) +2\alpha_x\beta_x
\end{align*}

For many coherent states from a distribution with variance $2 {\sigma^*}^2$ we get
\begin{align*}
    \left\langle \mathcal{E} \right\rangle(\epsilon)_x = \left\langle \left(\frac{\epsilon}{\sqrt{2}} (\hat{x}_a + \hat{x}_A - \hat{x}_b - \hat{x}_B) - (\alpha_x - \beta_x)\right)^2 \right\rangle = \\
    \frac{\epsilon^2}{2} \langle (\hat{x}_A-\hat{x}_B)^2 \rangle + \frac{\epsilon^2}{2} + {\sigma^*}^2 (\epsilon-1)^2 \,,
\end{align*}
 where we used that the distributions are centered around zero so $\mathrm{mean}(\alpha_x) = \mathrm{mean}(\beta_x)=0$. 

For $\left\langle \mathcal{E} \right\rangle(\epsilon)_p$ we proceed in a similar way, so in total we get: 
\begin{align}
    \left\langle \mathcal{E} \right\rangle(\epsilon) = \left\langle \mathcal{E} \right\rangle(\epsilon, r)_x + \left\langle \mathcal{E} \right\rangle(\epsilon)_p  = \frac{\epsilon^2}{2} (\langle (\hat{x}_A-\hat{x}_B)^2 + (\hat{p}_A+\hat{p}_B)^2 \rangle) + \epsilon^2 + 2 {\sigma^*}^2 (\epsilon-1)^2 
\end{align}

If Eve has access to a TMSV state with squeezing $\left\langle (\hat{x}_A-\hat{x}_B)^2\right\rangle + \left\langle (\hat{p}_A+\hat{p}_B)^2 \right\rangle = 2\ee^{-2r} < 2$ for $r>0$, the error becomes
\begin{align*}
   \left\langle \mathcal{E} \right\rangle(\epsilon, r) =   \epsilon^2 \ee^{-2r} + \epsilon^2 + 2 (\epsilon-1)^2 {\sigma^*}^2   
\end{align*}
For $r=0$, the state is separable, and Eve can only saturate the MDI bound without violating it.

To find Eve's optimal rescaling factor $\epsilon_{opt}$, we minimize the error: 
\begin{align*}
    \frac{\dd}{\dd \epsilon} \left\langle \mathcal{E} \right\rangle(\epsilon, r) = 2 \epsilon \ee^{-2r} + 2\epsilon + 4 (\epsilon-1) {\sigma^*}^2  =0 
\end{align*}
yielding
\begin{align*}
\epsilon_{opt} = \frac{2 {\sigma^*}^2}{2 {\sigma^*}^2 + \ee^{-2r} +1}
\end{align*}

For a separable state ($r = 0$), the resulting error is: 
\begin{align*}
    \left\langle \mathcal{E} \right\rangle(\epsilon_{opt}, 0) = 2\epsilon_{opt}^2 + 2 (\epsilon_{opt}-1)^2 {\sigma^*}^2 = \frac{2{\sigma^*}^2}{{\sigma^*}^2+1}
\end{align*}
which matches the MDI Bound $2v(\sigma^*,\sigma^*)$, ensuring Eve cannot falsely detect entanglement.

\begin{figure}[ht!]
    \centering
    \includegraphics[width=0.7\linewidth]{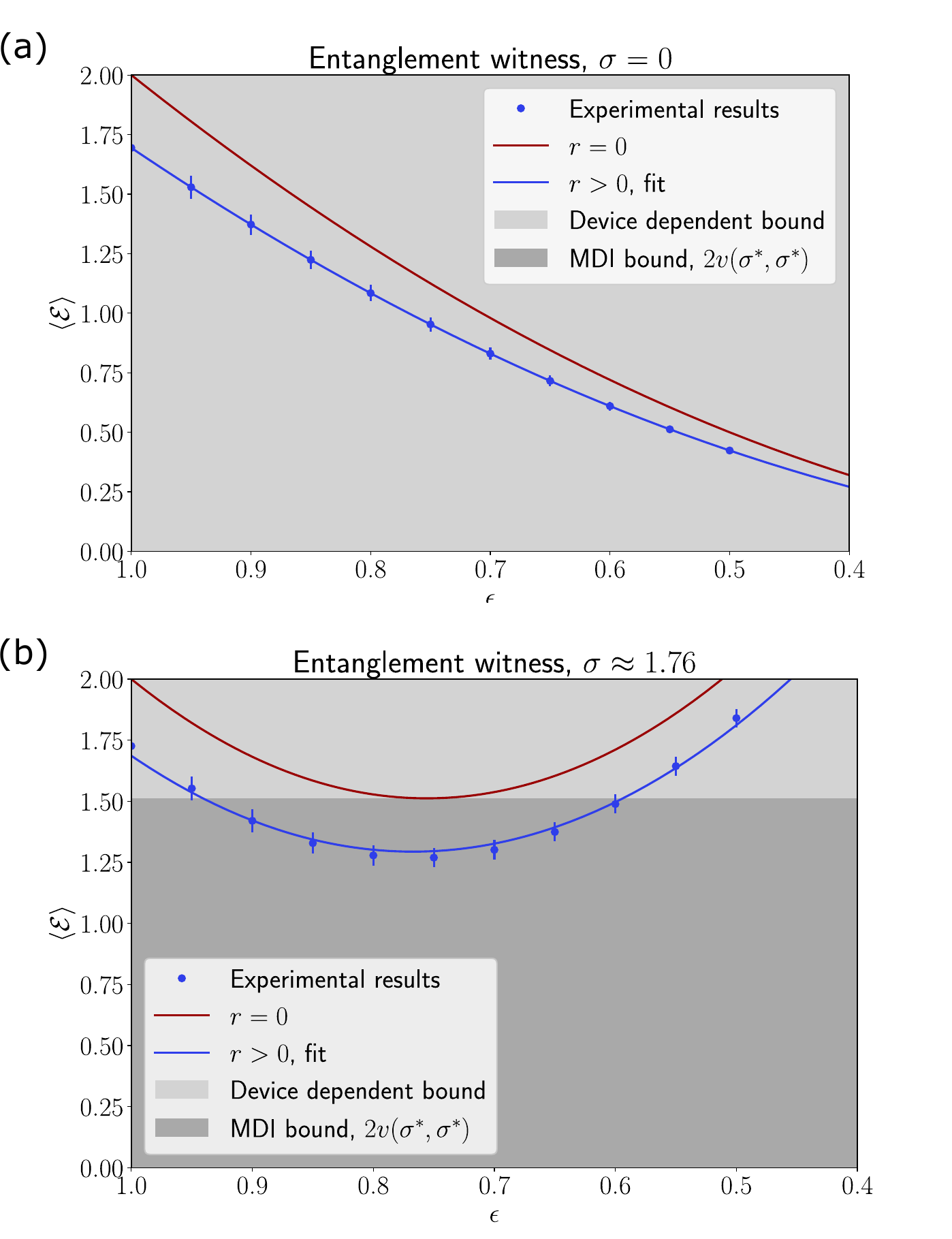}
    \caption{{\sf (a)} For separable states $(r=0)$, rescaling $\epsilon<1$ can lead to  false violations of the Simon-Duan criterion. {\sf (b)} For intermediate $\sigma$, rescaling by Eve saturates but does not violate the MDI bound. Entanglement $(r<1)$ allows the MDI bound to be violated as expected.}
    \label{fig:attack2}
\end{figure}

For Eve's rescaling, where the output is modified as $a_1 = \epsilon \frac{\hat{x}_a+\hat{x}_A}{\sqrt{2}}$, the predictions of the coherent states sent by Alice become inaccurate. This leads to an increase in the value of $\left\langle \mathcal{E} \right\rangle(\epsilon, r)$ when the measurements are rescaled. Thus, for lower values of $\epsilon$, the value of $\left\langle \mathcal{E} \right\rangle(\epsilon, r)$ increases. Eve's choice of $\epsilon$ involves a trade-off: reducing variance while maintaining the ability to correctly estimate the coherent states. This trade-off is evident in figure (4.a), which compares the effects of rescaling for a separable vacuum state ($r=0$, red line) and and entangled TMSV state ($r > 0$, blue line).

Without the trusted information held by Alice and Bob ($\sigma =0$), Eve can exploit rescaling to decrease the value of $\left\langle \mathcal{E} \right\rangle(\epsilon,r)$, potentially violating the device-dependent Simon-Duan bound with a separable state. However, the MDI bound $2 v(\sigma^*,\sigma^*)=0$ remains secure: even with an entangled TMSV state, it cannot be violated. This is depicted for varying $\epsilon$ in \cref{fig:attack2} {\sf(a)}. In \cref{fig:attack2} {\sf(b)}, we consider a scenario where Alice and Bob generate coherent states from a distribution with $\sigma^* \approx 1.76$. In this case, the value of $\left\langle \mathcal{E} \right\rangle(\epsilon,r = 0)$ decreases for small values of $\epsilon$. However, due to the trade-off between accurately predicting coherent states and reducing variance in Eve's homodyne measurements, the minimum value that Eve can achieve saturates the MDI bound but cannot violate it. For the case of non-zero squeezing ($r>0$), the MDI bound can be violated, as expected, due to the presence of both a non-separable state and additional information known only to Alice and Bob.



\begin{figure}
    \centering
    \includegraphics[width=\linewidth]{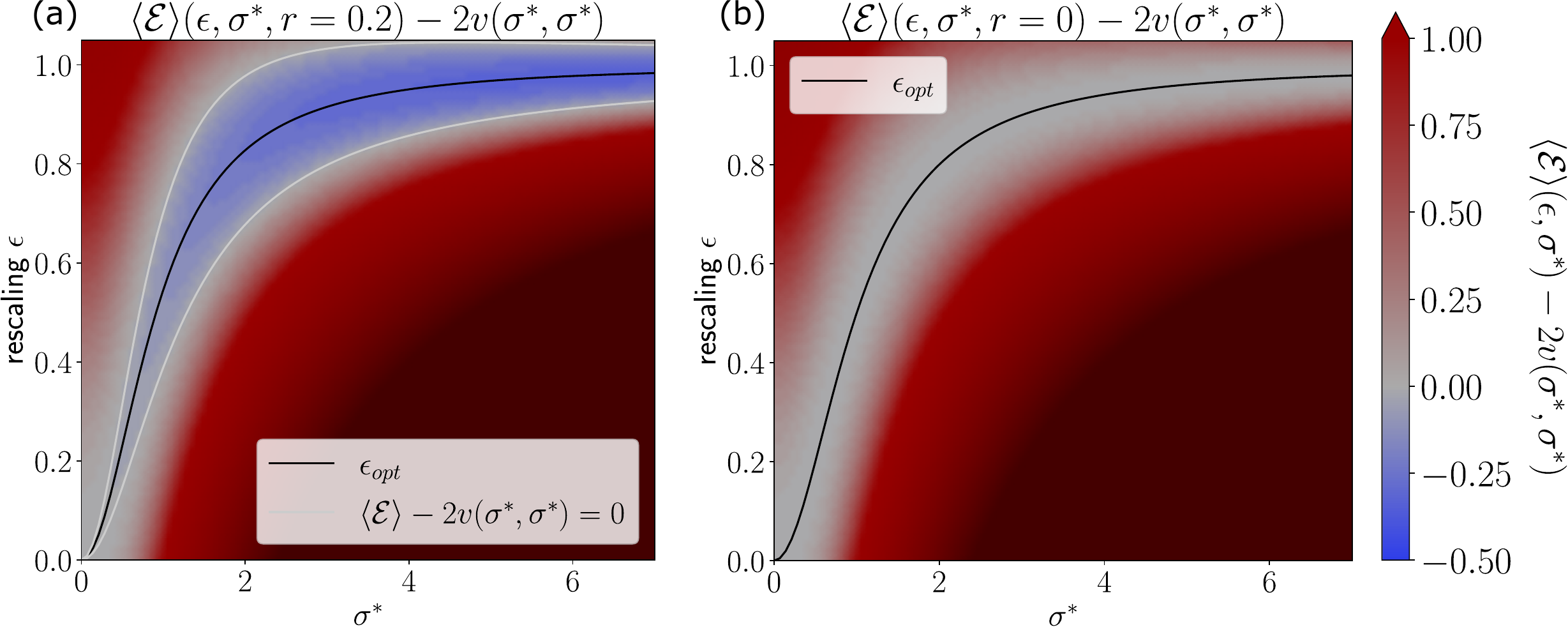}
    \caption{
    Contour plots of $\left\langle \mathcal{E} \right\rangle(\epsilon,r) - 2v(\sigma,\sigma)$ for {\sf (a)} $r=0.2$ and {\sf (b)} $r=0$. Regions where $\left\langle \mathcal{E} \right\rangle(\epsilon,r) - 2v(\sigma,\sigma)< 0$ indicate successful entanglement detection. For a separable state ($r=0$), as shown in  {\sf (b)}, the MDI bound cannot be violated even with shot noise rescaling ($\epsilon \neq 1$). However, for squeezed states ($r = 0.2$), as shown in {\sf (a)}, the bound can be violated, demonstrating the presence of entanglement.}
    \label{fig:attackContours}
\end{figure}

The regions in which the MDI bound can be violated, determined by $\left\langle \mathcal{E} \right\rangle(\epsilon,r) - 2v(\sigma,\sigma) \geq 0$, are shown in \cref{fig:attackContours}. 

\section{Memory witnessing}
If we model the memory as a lossy channel with transmission efficiency $\eta$ and excess noise $\xi$, the output quadrature after passing through the memory is given by: 
\begin{align*}
    \hat{x}'_a = \sqrt{\eta}  \hat{x}_a + \sqrt{1-\eta} \hat{x}_{00}.
\end{align*}
 where $\langle \Delta^2 \hat{x}_{00} \rangle = \frac{1+\xi}{2}$. To correctly estimate the input coherent state, Eve might need to amplify output of the channel. After amplification by a gain factor $\nu$, the output quadrature becomes: 
\begin{align*}
     \hat{x}''_a = \sqrt{\nu}  \hat{x}'_a + \sqrt{\nu-1} \hat{x}_{01} = \sqrt{\nu\eta} \hat{x}_a + \sqrt{\nu (1-\eta)}  \hat{x}_{00} + \sqrt{\nu-1} \hat{x}_{01},
\end{align*}

To match the state $\ket{\beta_{t = \tau}}$ with the state $\ket{\alpha_{t=0}}$, a similar scaling is applied: 
\begin{align*}
    \hat{x}'_b = \sqrt{\eta\nu}  \hat{x}_b + \sqrt{1-\eta\nu} \hat{x}_{02}.
\end{align*}
Eve then estimates the coherent state by performing homodyne detection and scaling the measurement outcome by $\frac{1}{\sqrt{\nu\eta}}$: 
\begin{align*}
    \hat{a} = \frac{\hat{x}''_a + \hat{x}'_b}{\sqrt{2\nu\eta}},
\end{align*}
and similarly for the $\hat{p}$-quadrature. 

The memory witness is evaluated as follows: 
\begin{align}
    \langle \mathrm{MDIEP} \rangle &= \langle (\hat{a} - \alpha_x - \beta_x)^2\rangle  + \langle (\hat{b}_p - \alpha_p + \beta_p)^2\rangle \nonumber \\
    &= \langle \hat{a}^2 \rangle - 2 \langle \hat{a} \rangle \underbrace{(\alpha_x + \beta_x)}_{= \langle \hat{a} \rangle } + (\alpha_x + \beta_x)^2 + \langle (\hat{b} - \alpha_p + \beta_p)^2\rangle \nonumber \\
    &= \langle \hat{a}^2 \rangle - \langle \hat{a} \rangle^2 + \langle \hat{b}^2 \rangle -\langle \hat{b} \rangle^2 = \langle \Delta^2 \hat{a} \rangle + \langle \Delta^2 \hat{b} \rangle,
\end{align}
We can now evaluate the variance of $\hat{a}$:
\begin{align*}
    \langle \Delta^2 \hat{a} \rangle &= \frac{1}{2\nu\eta} \langle \hat{x}''_a - \hat{x}'_b \rangle ^2  \\
    &= \frac{1}{2\nu\eta} (\eta \nu \langle \Delta^2 \hat{x}_a \rangle + \nu (1-\eta) \langle \Delta^2 \hat{x}_{00}\rangle \\
    &+ (\nu-1) \langle \Delta^2 \hat{x}_{01}\rangle + \nu\eta \langle \Delta^2 \hat{x}_p\rangle + (1-\nu\eta) \langle \Delta^2 \hat{x}_{02}\rangle) \\
    & = \frac{1}{4\nu\eta} (\eta \nu + \nu(1-\eta)(1+\xi)+ \nu-1 + \eta\nu+ 1 - \eta \nu) \\
    & = \frac{1}{4\eta} (2+\xi + \eta\xi).
\end{align*}
and similarly, for $\langle \Delta^2 \hat{b} \rangle$. By combining these results, we find: 
\begin{align}
     \langle \mathrm{MDIEP} \rangle = \frac{1}{2\eta} (2+\xi + \eta\xi) ,
\end{align}
which reduces to  $\langle \mathrm{MDIEP} \rangle = \frac{1}{\eta}$ for $\xi=0$. 

\begin{figure}
    \centering
    \includegraphics[width=\linewidth]{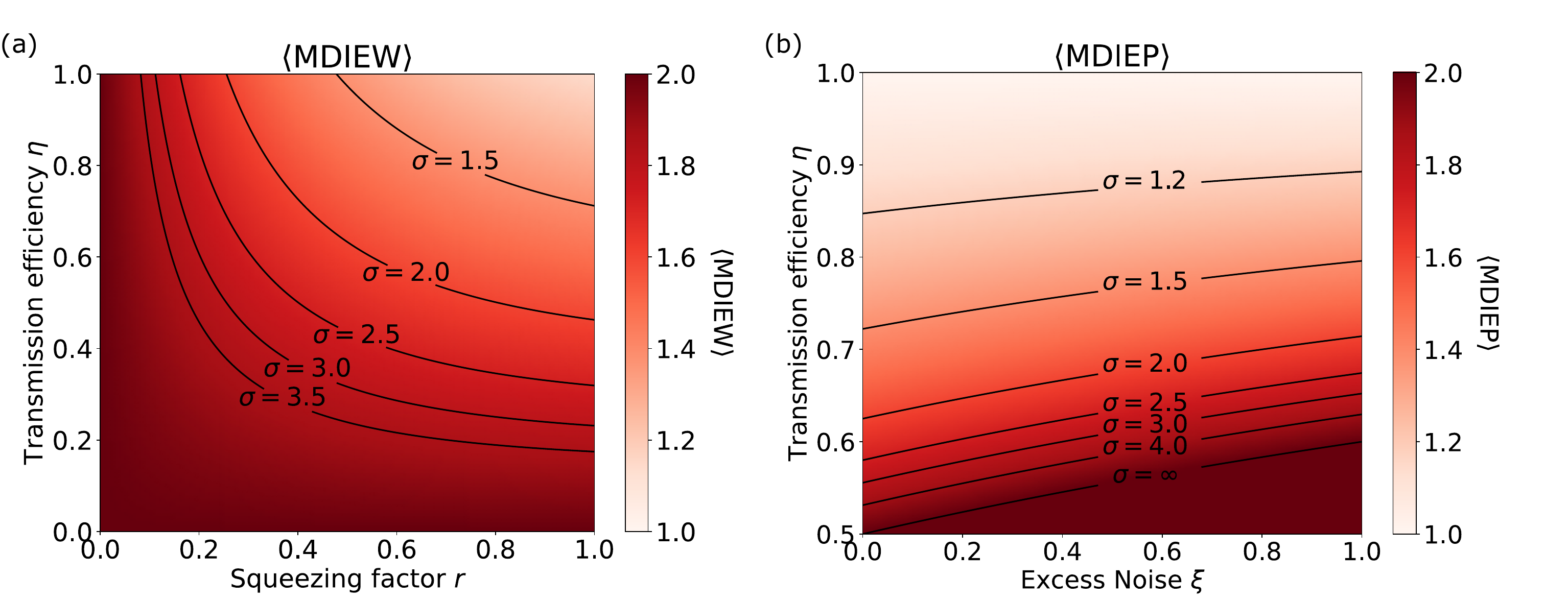}
    \caption{{\sf (a)} 
    Contour plot showing the values of $\langle \mathrm{MDIEW} \rangle$ achievable for different squeezing levels $r$ and transmission efficiencies $\eta$. The contours represent the values of $\sigma$ required to violate the bound $\langle \mathrm{MDIEW} \rangle \geq 2 v(\sigma,\sigma)$ 
    {\sf (b)}
    Contour plot of the values of $\langle \mathrm{MDIEP} \rangle$ achievable for different values of transmission efficiency $\eta$ and excess noise $\xi$. The contours indicate the values of $\sigma$ needed to violate the bound  $\langle \mathrm{MDIEP} \rangle \geq 2 v(\sigma,\sigma)$ 
    }
    \label{fig:Countous}
\end{figure}

The witness does not depend on the gain factor $\nu$, performed in postprocessing. The values of $ \langle \mathrm{MDIEP} \rangle$ for varying $\eta$ and $\xi$ are shown in \cref{fig:Countous} {\sf (b)}

\section{Phase noise model}

We now examine the impact of phase noise in the experimental setup. Three types of phase noise need to be considered: 1) Phase between the to SMSV states during the preparation of the TMSV state. 2) Phase noise between the TMSV and the detectors. 3) Phase noise between the coherent states and the detectors. To simplify the analysis, we assume that the phase noise is consistent across both modes of the TMSV and all four detectors, as well as across both coherent states and all four detectors. 

We model phase noise as random phase shifts. A phase shift $\theta_1$ represents the phase noise between the TMSV and the homodyne detectors, while $\theta_2$ represents the phase noise between the coherent states and the detectors. The quadrature transformations are expressed as 
\begin{align*}
    \hat{x}_A \rightarrow \cos(\theta_1) \hat{x}_A + \sin(\theta_1) \hat{p}_A \\
    \hat{p}_A \rightarrow \cos(\theta_1) \hat{p}_A - \sin(\theta_1) \hat{x}_A \\
    \hat{x}_a \rightarrow \cos(\theta_2) \hat{x}_a + \sin(\theta_2) \hat{p}_a \\
    \hat{p}_a \rightarrow \cos(\theta_2) \hat{p}_a - \sin(\theta_2) \hat{x}_a 
\end{align*}
and similarly for $\hat{x}_B,\hat{p}_B,\hat{x}_b,\hat{p}_b$. Additionally, We account for transmission losses, assuming for simplicity that $\eta_A = \eta_B$.

Next, we consider a phase shift of $\theta_3$ between the two modes of the SMSV, leading to the following transformation: 
\begin{align*}
    \langle \hat{x}_A^2 \rangle &= \cosh(2r)/2 \rightarrow \langle \hat{x}_A^2 \rangle = \cosh(2r)/2 \\
    \langle \hat{x}_B^2 \rangle &= \langle \hat{p}_A^2 \rangle= \langle \hat{p}_B^2\rangle \\
    \langle \hat{x}_A \hat{x}_B\rangle &= \sinh(2r) /2 \rightarrow  \langle \hat{x}_A \hat{x}_B\rangle = \cos(\theta_3) \sinh(2r) /2 \\
    \langle \hat{x}_A \hat{x}_B\rangle &= - \langle \hat{p}_A \hat{p}_B\rangle \\
    \langle \hat{x}_A \hat{p}_B\rangle &= 0 \rightarrow  \langle \hat{x}_A \hat{p}_B\rangle = -\sin(\theta_3) \sinh(2r) /2 \\
    \langle \hat{p}_A \hat{x}_B\rangle &= \langle \hat{x}_A \hat{p}_B\rangle
\end{align*}

Substituting these transformations into the error expression, we find:
\begin{align*}
    \langle \mathcal{E} \rangle(\epsilon,r,\eta, \theta_1, \theta_2, \theta_3)
    = \epsilon^2 \eta \Big( \cos^2(\theta_1) (\cosh(2r)-\cos(\theta_3)\sinh(2r))-\sin^2(\theta_1) (\cosh(2r)+cos(\theta_3)\sinh(2r)) \\
    +2\cos(\theta_1)\sin(\theta_1)\sin(\theta_3) \sinh(2r) \Big) + (1-\eta)\epsilon^2 + \epsilon^2 + 2\sigma^2 (1+\epsilon^2 - 2\epsilon \cos(\theta_2))
\end{align*}
For phase noise modeled as Gaussian-distributed random variables with variances $\Delta^2 \theta_1,\Delta^2 \theta_2$ and $\Delta^2 \theta_3$, the average becomes~\citep{suleiman202240}
\begin{align*}
    &\langle \mathcal{E} \rangle(\epsilon,r,\eta, \Delta^2 \theta_1, \Delta^2 \theta_2, \Delta^2 \theta_3) \\
    =&\int_{-\infty}^\infty \dd \theta_1 \dd \theta_2 \dd \theta_3 
    \frac{1}{\sqrt{2\pi\Delta^2 \theta_1}} \frac{1}{\sqrt{2\pi\Delta^2 \theta_2}} \frac{1}{\sqrt{2\pi\Delta^2 \theta_3}} 
    \ee^{-\theta_1^2 /(2\Delta^2 \theta_1)}
    \ee^{-\theta_2^2 /(2\Delta^2 \theta_2)}
    \ee^{-\theta_3^2 /(2\Delta^2 \theta_3)}
    \langle \mathrm{MDIEW} \rangle(\epsilon,r,\eta, \theta_1, \theta_2, \theta_3)\\
    =& \epsilon^2 \eta \Biggl[ 
    \frac{1+\ee^{-2\Delta^2 \theta_1}}{2} \left\{ \cosh(2r)- \ee^{-\Delta^2\theta_3/2}\sinh(2r) \right\} \\
    &+\frac{1-\ee^{-2\Delta^2 \theta_1}}{2} \left\{ \cosh(2r)+ \ee^{-\Delta^2\theta_3/2}\sinh(2r) \right\} \Biggr] \\
    &+ (1-\eta)\epsilon^2 + \epsilon^2 + 2\sigma^2 \left( 1 + \epsilon^2 - 2\epsilon \ee^{-\Delta^2\theta_2/2} \right)
\end{align*}



Without phase noise and assuming $\epsilon =1$, the error reduces to 
\begin{align*}
    \langle \mathcal{E} \rangle(1,r,\eta, 0, 0, 0)  = \langle \mathrm{MDIEW} \rangle =  2 + \eta (\ee^{-2r}-1).
\end{align*}
The corresponding values can be visualized in \cref{fig:Countous} {\sf (a)}. With phase noise, the error includes the effects of phase noise, leading to a decrease in the observed squeezing and consequently affecting the ability to witness entanglement. 

\end{document}